\documentclass[journal]{IEEEtran}

\usepackage{graphicx}
\usepackage{enumerate}
\usepackage{cite}
\usepackage{bm}
\usepackage[dvipsnames]{xcolor}
\usepackage{amsmath,amssymb,amsthm}
\usepackage{dutchcal}
\usepackage{stfloats}
\usepackage{suffix}
\usepackage{mathtools}
\usepackage[hidelinks]{hyperref}
\begin{document}

\title{Dual-Polarized Massive MIMO-RSMA Networks: Tackling Imperfect SIC}

\author{Arthur Sousa de Sena, \textit{Member}, \textit{IEEE}, Pedro H. J. Nardelli,  \textit{Senior Member}, \textit{IEEE}, Daniel Benevides da Costa, \textit{Senior Member}, \textit{IEEE}, Petar Popovski, \textit{Fellow}, \textit{IEEE}, Constantinos B. Papadias, \textit{Fellow}, \textit{IEEE}, and Mérouane Debbah, \textit{Fellow}, \textit{IEEE}
\thanks{This work was partly supported by the Academy of Finland via: (a) FIREMAN consortium n.326270 as part of CHIST-ERA grant CHIST-ERA-17-BDSI-003,  (b) EnergyNet Fellowship n.321265/n.328869/n.352654, and (c) X-SDEN project n.349965. This work was also partly supported by the Villum Investigator Grant ``WATER'' from the Velux Foundation, Denmark. Part of this article has been accepted for presentation at the IEEE Global Communications Conference (GLOBECOM), Rio de Janeiro, Brazil, December 2022. (\it Corresponding authors: Arthur Sousa de Sena and Daniel Benevides da Costa.)}
\thanks{Arthur Sousa de Sena is with the Technology Innovation Institute, 9639 Masdar City, Abu Dhabi, the United Arab Emirates, and with Lappeenranta-Lahti University of Technology, 53850 Lappeenranta, Finland (email: arthurssena@ieee.org).}
% %
\thanks{Pedro H. J. Nardelli is with Lappeenranta-Lahti University of Technology, 53850 Lappeenranta, Finland (email: pedro.nardelli@lut.fi).}
% %
\thanks{Daniel Benevides da Costa and Mérouane Debbah are with the Technology Innovation Institute, 9639 Masdar City, Abu Dhabi, the United Arab Emirates (email: danielbcosta@ieee.org, merouane.debbah@tii.ae).}
% %
\thanks{Petar Popovski is with Aalborg University, 9220 Aalborg, Denmark (email: petarp@es.aau.dk).}
% %
\thanks{Constantinos B. Papadias is with The American College of Greece, 15342 Athens, Greece (email: cpapadias@acg.edu).}
% %
}

\maketitle

\begin{abstract}
    The polarization domain provides an extra degree of freedom (DoF) for improving the performance of multiple-input multiple-output (MIMO) systems. This paper takes advantage of this additional DoF to alleviate practical issues of successive interference cancellation (SIC) in rate-splitting multiple access (RSMA) schemes. Specifically, we propose three dual-polarized downlink transmission approaches for a massive MIMO-RSMA network under the effects of polarization interference and residual errors of imperfect SIC. The first approach implements polarization multiplexing for transmitting the users' data messages, which removes the need to execute SIC in the reception. The second approach transmits replicas of users' messages in the two polarizations, which enables users to exploit diversity through the polarization domain.
    The third approach, in its turn, employs the original SIC-based RSMA technique per polarization, and this allows the BS to transmit two independent superimposed data streams simultaneously. An in-depth theoretical analysis is carried out, in which we derive tight closed-form approximations for the outage probabilities of the three proposed approaches. Accurate approximations for the ergodic sum-rates of the two first schemes are also derived. Simulation results validate the theoretical analysis and confirm the effectiveness of the proposed schemes. For instance, under low to moderate cross-polar interference, the results show that, even under high levels of residual SIC error, our dual-polarized MIMO-RSMA strategies outperform the conventional single-polarized MIMO-RSMA counterpart. It is also shown that the performance of all RSMA schemes is impressively higher than that of single and dual-polarized massive MIMO systems employing non-orthogonal multiple access (NOMA) and orthogonal multiple access (OMA) techniques. 
\end{abstract}


\IEEEpeerreviewmaketitle

\section{Introduction}
Massive multiple-input multiple-output (MIMO) has become an indispensable technology for fifth-generation (5G) wireless communications systems and beyond. With a large number of antennas at the base station (BS), massive MIMO can serve multiple users simultaneously and achieve near-optimal performance through low-complexity linear precoding techniques \cite{ni3, ref1, Liu20, Li21}. This capability gives the massive MIMO technology remarkable spectrum and connectivity advantages over small-scale MIMO and single-antenna systems. Nevertheless, the installation of tens to hundreds of antenna elements in a tight physical space can create a strong correlation among adjacent antennas. This issue puts hard constraints on the size of practical antenna arrays and can hamper the performance of massive MIMO \cite{ref1}. Fortunately, the polarization domain provides an efficient way to mitigate this limitation. Specifically, since electromagnetic waves with orthogonal polarizations propagate with a low correlation, it is possible to implement orthogonal dual-polarized antennas organized into co-located pairs and build an array with twice the number of antennas of a single-polarized array using identical physical dimensions \cite{ni3}. In addition to the space efficiency, the polarization domain offers a new degree of freedom (DoF) to MIMO systems, which can be exploited for generating both multiplexing and diversity gains. These advantages bring to dual-polarized MIMO systems performance improvements that can remarkably outperform single-polarized counterparts \cite{ni3,ref1, SenaL21, SenaT21}.

Efficient multiple access (MA) techniques are also essential for supporting the stringent connectivity and spectrum requirements expected in beyond-5G systems. In particular, rate-splitting multiple access (RSMA) has arisen as a robust next-generation MA technique for MIMO systems with the potential for appealing performance improvements \cite{Yang13, Clerckx16, Mao18, Dizdar21}. 
RSMA can be seen as an extension of the concept of rate splitting from \cite{Carleial78}, which has been proposed as an attempt to augment the region of achievable rates in two-user broadcast interference channels. The author of \cite{Carleial78} revealed that if users undergo interfering channels, it can be beneficial to transmit part of the data messages via a shared common stream and decode part of the inter-user interference \cite{SenaM22}. The modern RSMA technique generalizes and exploits this feature to overcome residual interference in broadcast MIMO channels, which conventional precoding schemes cannot cancel, e.g., due to imperfect channel state information (CSI).
%RSMA implements the concept of rate splitting from \cite{Carleial78}, which has been proposed as an attempt to augment the region of achievable rates.
At the BS, the RSMA technique divides each users' message into two parts. The first part of each message is encoded into a single super symbol, called the common message, and mapped to the BS antennas through a common precoder, which is intended for all uses. The second part of each message, called private message, is transmitted via a private precoder that should be decoded only at the intended user. Upon reception, users rely on successive interference cancellation (SIC) to recover the transmitted messages. This strategy enables RSMA to flexibly manage the amount of interference that is decoded and the amount that is treated as noise. Thanks to these features, RSMA can deliver high spectral \cite{Joudeh16} and energy efficiencies \cite{MaoC18}, optimality in terms of DoF \cite{Piovano17}, and robustness even in scenarios with imperfect CSI \cite{Mao18, Joudeh16, MaoC18, Piovano17}.
Moreover, recent studies show that RSMA can boost the performance and mitigate practical issues of massive MIMO systems, such as pilot contamination \cite{Thomas20}, hardware imperfection problems \cite{Papazafeiropoulos17}, and issues of high mobility \cite{Dizdar21}. These capabilities enable RSMA to outperform all conventional MA techniques, including space-division multiple access (SDMA), non-orthogonal multiple access (NOMA), and orthogonal multiple access (OMA) techniques like time-division multiple access (TDMA) and orthogonal frequency-division multiple access (OFDMA) \cite{Dizdar21,Thomas20,Papazafeiropoulos17,Dai16}.

Despite the above benefits, there are still issues that need to be studied and tackled before massive MIMO-RSMA systems become commercially available. As mentioned before, RSMA relies on SIC to recover the transmitted messages. Even though SIC introduces advantages, it also has its limitations. First, due to the SIC protocol in RSMA, the common message is always detected with interference from private messages, which has degrading effects on the system data rates.
Moreover, executing SIC perfectly in practical systems is a difficult task. Due to hardware limitations, degraded CSI, and other issues, SIC errors are likely to happen in practice. As a result, the private messages are inevitably recovered with residual interference from the common message.
As demonstrated in \cite{SenaISIC2020}, the residual interference left by imperfect SIC can strongly harm the performance of SIC-based schemes. 
%, which implies that SIC issues are also detrimental to the performance of MIMO-RSMA.
This fact implies that strategies for combatting the effects of imperfect SIC in RSMA are necessary. In particular, the performance superiority and additional DoF of dual-polarized MIMO systems can be exploited to alleviate interference issues of SIC \cite{SenaT21}. Nevertheless, as discussed in the following subsection, the study of dual-polarized MIMO-RSMA systems are still missing in the literature.

\subsection{Related Works}

Existing MIMO-RSMA-related works consider only single-polarized system models. 
For instance, the concept of single-layer RSMA, which relies on a single common message, was first studied in \cite{Yang13} considering single-polarized MIMO broadcast channels. The authors investigated the DoF region achieved by the RSMA approach in a two-user scenario considering imperfect and delayed CSI. The proposed scheme outperformed baseline systems both in terms of DoF and sum-rate. The advantages of single-layer RSMA were further investigated in \cite{Clerckx16}, also considering two-user single-polarized MIMO scenarios. This work demonstrated that MIMO-RSMA is way more robust to imperfect CSI than conventional MIMO-SDMA and MIMO-TDMA schemes. The authors of \cite{Dai16} extended the RSMA concept and proposed a two-layer hierarchical rate-splitting (HRS) strategy for a single-polarized multi-group massive MIMO network with multiple spatially correlated single-antenna users. This work demonstrated that when different groups of users experience overlapping eigenspaces, it is advantageous to convey two layers of common messages.
The work in \cite{Mao18} also studied a MIMO network with multiple single-antenna users and showed that MIMO-RSMA substantially outperforms both conventional multi-user MIMO and MIMO-NOMA schemes in underloaded and overloaded scenarios.
Mobility issues in single-polarized massive MIMO-RSMA networks were studied in \cite{Dizdar21}. Specifically, lower bounds for the ergodic sum-rate of MIMO-RSMA were derived. The power allocation between common and private messages was also studied. %This work demonstrated with link-level simulations that the proposed massive MIMO-RSMA strategy is significantly more robust to the effects of user mobility than conventional multi-user massive MIMO systems. 
The work in \cite{Mao20} studied the max-min fairness of MIMO-RSMA in cooperative user relaying networks, where an algorithm based on successive convex approximation was proposed to jointly optimize the precoders, message split, and time slot allocation. In \cite{Yin21}, MIMO-RSMA was investigated in multi-beam satellite systems. 
%The authors characterized the system DoF and developed a max-min transmit power optimization strategy for rate fairness.
%The proposed single-polarized MIMO-RSMA satellite scheme achieved attractive performance gains. 
The combination of RSMA and visible-light communications in a two-user scenario was investigated in \cite{Naser20}. In \cite{Ahmad21}, the authors studied the benefits of MIMO-RSMA in a cloud radio access network, and the multicarrier MIMO-RSMA case was addressed in \cite{Li20}. For a more complete literature review on the RSMA subject, readers are referred to the work in \cite{Mao22}.

Works studying dual-polarized MIMO systems have also appeared in parallel developments recently. For example, the authors of \cite{ni3} proposed polarization diversity and multiplexing strategies for a dual-polarized MIMO-NOMA network. The proposed dual-polarized schemes significantly outperformed conventional single-polarized counterparts in the presented simulation results. In \cite{Khalilsarai22}, the authors proposed a novel channel estimation framework for dual-polarized MIMO systems operating in frequency division duplexing (FDD) mode. Hybrid beamforming strategies for dual-polarized millimeter-wave MIMO systems were proposed in \cite{Sung18} and \cite{Kim20}. The work in \cite{Park20} investigated the performance of a massive MIMO system with distributed polarized antennas, and dual-polarized MIMO systems assisted by intelligent reflecting surfaces (IRSs) were considered in \cite{SenaL21, SenaT21}, and \cite{Chen21}.

\subsection{Motivation and Contributions}
Even though there are some RSMA-related works that model imperfect SIC \cite{Dizdar20, Jihye21, Mishra22}, the harmful effects of this major problem are still not totally clarified. Furthermore, even though dual-polarized MIMO systems with different MA techniques have been investigated in recent works, this paper is the first in the literature to propose a dual-polarized MIMO-RSMA network and study ways to mitigate interference issues of SIC via the polarization domain. Only \cite{SenaL21} and \cite{SenaT21} have harnessed the features of wave polarization to alleviate SIC issues. However, these works studied dual-polarized MIMO-NOMA schemes assisted by IRSs, which are system models fundamentally different from the one proposed in this paper. This major gap in the literature motivates the development of this work. Further details and our main contributions are summarized as follows:
\begin{itemize}
    \item By modeling the practical issues of imperfect SIC and depolarization phenomena, we propose a two-stage dual-polarized MIMO-RSMA transmission scheme.
    In the first stage, inspired by the precoding strategy for dual-polarized channels from \cite{ref1}, we design dual-polarized outer precoders for performing spatial multiplexing of multiple groups of spatially correlated users located in different angular sectors. To this end, we exploit only the channel covariance information of the users within each group. This dual-polarized strategy can be seen as an extension of the single-polarized precoders designed for spatially correlated massive MIMO scenarios in \cite{Dai16} and \cite{ref6}. Nevertheless, our work is the first to apply the concept to dual-polarized RSMA-based schemes.
    In the second stage, we propose private and common precoders for implementing three low-complexity dual-polarized RSMA strategies. These inner precoders are designed based on the instantaneous realizations of reduced-dimension effective channel gains, which demand a small feedback overhead.
    
    \item The first proposed dual-polarized MIMO-RSMA strategy implements polarization multiplexing to transmit common and private messages in parallel, which removes the need to execute SIC in the receivers. At the cost of introducing polarization interference, this approach can free the system from the detrimental effects of imperfect SIC and enable all users to experience a reduced overall interference when decoding the desired messages.
    
    \item The second dual-polarized MIMO-RSMA strategy transmits replicas of the common and private messages in the two polarizations and enables users to achieve diversity in the polarization domain. Like the original RSMA scheme, this approach also relies on SIC to decode users' messages in the reception. However, polarization diversity alleviates the degradation caused by SIC interference, which contributes to substantial gains in terms of outage probability.
    
    \item The third dual-polarized MIMO-RSMA strategy employs the original SIC-based RSMA technique per polarization, which allows the BS to transmit two independent superimposed data streams simultaneously. As a result, in ideal scenarios with perfect SIC decoding, this third strategy offers users significant sum-rate gains over the two first approaches. However, the scheme is the least reliable, i.e., it suffers from the highest outage probabilities among the proposed schemes.
    
    \item We carry out an in-depth theoretical study on the proposed transmission approaches. We perform, first, a statistical characterization on the effective channel gains observed in the common and private data streams, which turns out to be correlated. As a result, the derivation of the exact distributions for the signal-to-interference-plus-noise ratios (SINRs) becomes challenging. Alternatively, we assume that the SINR gains are statistically independent and determine approximate distributions. Closed-form expressions for the outage probabilities of the three MA strategies are derived based on the obtained distributions. Last, we provide tight approximations for the ergodic sum-rates experienced with the first two schemes. % experienced with the two proposed approaches.
    
    \item Simulation and numerical results supported by comprehensive discussions validate the theoretical analysis and provide important insights into the performance of the proposed schemes, such as the associated trade-offs between throughput and reliability. Our results show that the first proposed scheme is advantageous mainly in scenarios with high SIC interference, whereas the second dual-polarized RSMA strategy with polarization diversity achieves the lowest outage probabilities particularly when the SIC decoding errors are low to moderate. In contrast, the third scheme based on the original RSMA achieves the worst outage probabilities but delivers the highest sum-rates if the levels of SIC error are not excessively high. Moreover, aligned with recent works which demonstrate that RSMA outperforms SDMA, NOMA, and OMA in single-polarized MIMO systems \cite{Mao18}, our simulation results reassure these performance gains and show that RSMA schemes can achieve data rates impressively higher than those baseline techniques also in dual-polarized MIMO systems, even in scenarios with imperfect SIC and polarization interference.
\end{itemize}

\noindent  \textbf{\textit{Notation and Special Functions:}} Bold-faced lower-case letters denote vectors and upper-case represent matrices. The transpose and the Hermitian transpose of $\mathbf{A}$ are represented, respectively, by $\mathbf{A}^T$ and $\mathbf{A}^H$, the operator tr$\{\mathbf{A}\}$ computes the trace of $\mathbf{A}$, and $[\mathbf{A}]_{i:j}$ returns a sub-matrix of $\mathbf{A}$ containing its columns from $i$ to $j$. The symbol $\otimes$ represents the Kronecker product, $\mathbf{I}_M$ represents the identity matrix of dimension $M\times M$, and $\mathbf{0}_{M, N}$ denotes the $M\times N$ matrix with all zero entries. In addition, $\mathrm{E}(\cdot)$ denotes expectation, $\Gamma(\cdot)$ is the Gamma function \cite[eq. (8.310.1)]{ref8}, $\gamma(\cdot,\cdot)$ is the lower incomplete Gamma function \cite[eq. (8.350.1)]{ref8}, $\bm{e}_n(\cdot)$ denotes the truncated Taylor series of the exponential function with $n$ terms \cite[eq. (1.211.1)]{ref8}, and $\mathrm{Ei}(\cdot)$ corresponds to the exponential integral \cite[eq. (8.211.1)]{ref8}.

\section{System Model}

We consider a massive MIMO network in which a single BS communicates in downlink mode with $L$ users. The BS is equipped with $M/2$ pairs of co-located dual-polarized transmit antennas, and each user employs one pair of dual-polarized receive antennas, such that each antenna pair contains one horizontally and one vertically polarized antenna element. Moreover, the antenna pairs at the BS are uniformly spaced by $\lambda/2$, in which $\lambda$ denotes the carrier wavelength and $M\gg 1$. Due to the scattering environment and the closely spaced antennas, the wireless channels of users located in similar angular sectors become correlated. The BS exploits this characteristic and clusters the users into $G$ groups based on the likeness of their channel covariance matrices. For simplicity, we assume that each group contains $U$ users\footnote{The dual-polarized MIMO-RSMA schemes proposed in this work are also applicable to the case where distinct groups comprise different numbers of users. However, the study of these more general scenarios is left for future work.}, i.e., $L = G U$, and that users within each group share a common covariance matrix $\mathbf{\bar{R}}_g = \mathbf{I}_{2} \otimes \mathbf{R}_g \in \mathbb{C}^{M\times M}$, in which $\mathbf{R}_g \in \mathbb{C}^{\frac{M}{2} \times \frac{M}{2}}$ is the covariance matrix corresponding to each polarization with rank denoted by $r_g$. Under such assumptions, we can invoke the Karhunen-Loeve decomposition and represent the wireless channel for the $u$th user in the $g$th group by \cite{ni3}
\begin{align}\label{eq01}
    \mathbf{H}_{gu} &= \sqrt{\zeta_{gu}}\begin{bmatrix} \mathbf{h}^{vv}_{gu} & \sqrt{\chi}\mathbf{h}^{vh}_{gu} \\
    \sqrt{\chi}\mathbf{h}^{hv}_{gu} & \mathbf{h}^{hh}_{gu}\end{bmatrix} \nonumber\\
     &=  \begin{bmatrix} \sqrt{\zeta_{gu}}\mathbf{U}_g\bm{\Lambda}^{\frac{1}{2}}_g\mathbf{g}^{vv}_{gu} & \sqrt{\zeta_{gu} \chi}\mathbf{U}_g\bm{\Lambda}^{\frac{1}{2}}_g\mathbf{g}^{vh}_{gu} \\ 
    \sqrt{\zeta_{gu} \chi}\mathbf{U}_g\bm{\Lambda}^{\frac{1}{2}}_g \mathbf{g}^{hv}_{gu} & \sqrt{\zeta_{gu}}\mathbf{U}_g\bm{\Lambda}^{\frac{1}{2}}_g\mathbf{g}^{hh}_{gu}\end{bmatrix} \in \mathbb{C}^{M\times 2},
\end{align}
where $\bm{\Lambda}_g \in \mathbb{R}^{\bar{r}_g \times \bar{r}_g}_{>0}$ is a diagonal matrix formed by $\bar{r}_g$ nonzero eigenvalues of $\mathbf{R}_g$ sorted in descending order, $\mathbf{U}_g \in \mathbb{C}^{\frac{M}{2}\times \bar{r}_g}$ comprises the corresponding $\bar{r}_g$ left eigenvectors of $\mathbf{R}_g$ obtained from the singular value decomposition (SVD), $\mathbf{g}^{ij}_{gu} \in \mathbb{C}^{\bar{r}_g \times 1}$ is a vector that models the reduced-dimension fast-fading channels from polarization $i$ to polarization $j$, in which $i, j \in \{v, h\}$, with $v$ and $h$ denoting, respectively, the vertical and horizontal polarizations, whose entries follow the complex Gaussian distribution with zero-mean and unit-variance, $\zeta_{gu}$ denotes the large-scale fading coefficient, and $\chi$ represents the inverse of the cross-polar discrimination (iXPD) parameter\footnote{Experiments in a practical industrial environment in \cite{Challita20} achieved iXPD values of $\chi = 0.02$ and $\chi = 0.09$ for line-of-sight and non-line-of-sight scenarios, respectively. Recent dual-polarized antenna array prototypes revealed that these values can be further improved. A novel array architecture proposed in \cite{Yamada19}, for instance, has reached an iXPD as low as $\chi = 0.00025$.} that models the level of cross-polar interference experienced in the system, in which $\chi = 0$ corresponds to the ideal scenario with no interference, and $\chi = 1$ models the extreme case with maximum interference.

\begin{figure}[t]
	\centering
	\includegraphics[width=1\linewidth]{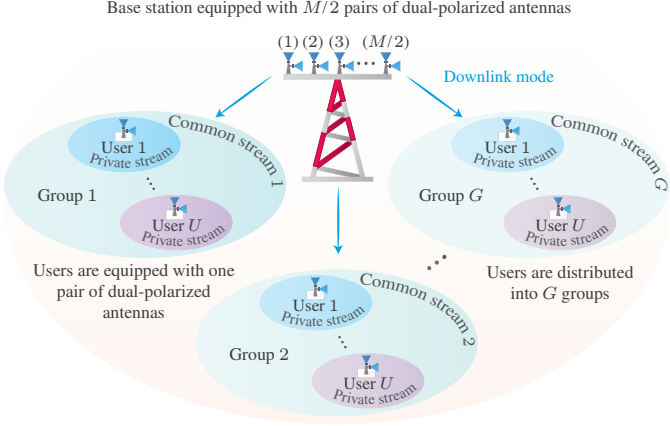}
	\caption{Proposed system model. A dual-polarized massive MIMO-RSMA base station serves multiple dual-polarized users distributed into different spatial groups.}\label{f1}
\end{figure}

\subsection{Rate-splitting multiple access for dual-polarized massive MIMO}
This subsection provides more details on the RSMA strategies proposed in this work. Specifically, inspired by the recent works \cite{ni3, ref1}, \cite{Dai16}, we implement a two-stage transmission approach where, in the first stage, spatial multiplexing is performed for separating the multiple groups of users and, in the second stage, the RSMA technique is employed to serve the users within each group. 
To this end, the BS splits the uncoded data message, denoted by $s_{gu}$, intended for the $u$th user within the $g$th group, into a common part, $s^{0}_{gu}$, and a private part, $s^{1}_{gu}$, such that $s_{gu} = (s^{0}_{gu}, s^{1}_{gu})$. The common parts of all users within the $g$th group are encoded into a single common symbol $c_g = (s^{0}_{g1}, \cdots, s^{0}_{gU})$, for $g=1, \cdots, G$, while the private parts of each individual user, $s^{1}_{gu}$, are encoded separately into private symbols $p_{gu}$. The symbol $c_g$ should be decoded by all users within the $g$th group, whereas $p_{gu}$ should be decoded by only its intended $u$th user. Last, $c_g$ and $p_{gu}$ are multiplied by linear precoders and then superimposed in the power domain for transmission, resulting in the following data stream:
\begin{align}\label{eqsig1}
    \mathbf{x} =\hspace{-1mm} \sum_{m = 1}^{G}  \mathbf{K}_{m}\left(\hspace{-.5mm} \mathbf{c}_{m} \sqrt{\alpha_{m}} c_{m} +  \sum_{n = 1}^{U} \mathbf{p}_{mn} \sqrt{\beta_{mn}} p_{mn} \hspace{-.5mm}\right)
    \in \mathbb{C}^{M \times 1},
\end{align}
where $\alpha_{m}$ and $\beta_{mn}$ denote, respectively, the power allocation coefficients for the common and private messages, $\mathbf{K}_{m}  = \mathbf{I}_{2} \otimes \mathbf{F}_m \in \mathbb{C}^{M\times\bar{M}}$ is the precoding matrix responsable for performing spatial multiplexing of user groups, in which $\mathbf{F}_m \in \mathbb{C}^{\frac{M}{2}\times\frac{\bar{M}}{2}}$ represents the precoding matrix for each polarization, $\bar{M}$ is a parameter that controls the dimension of the projected channel, and $\mathbf{c}_{m} \in \mathbb{C}^{\bar{M} \times 1}$ and $\mathbf{p}_{mn} \in \mathbb{C}^{\bar{M} \times 1}$ are the precoding vectors for the common and private messages, respectively, satisfying $\|\mathbf{c}_{m}\|^2 = 1$ and $\|\mathbf{p}_{mn}\|^2 = 1$.

At the users' side, the common symbol is detected first while treating the private symbols as interference. As a result, users achieve an estimate of $c_g$, denoted by $\hat{c}_g$. Then, each user executes SIC to subtract $\hat{c}_g$ from the composite signal in \eqref{eqsig1}. After that, an estimate of the private symbol, represented by $\hat{p}_{gu}$, can be obtained. Next, the $u$th user decodes the common symbol $\hat{c}_g$ and recovers the estimate of its common part $\hat{s}^{0}_{gu}$. Meanwhile, the private symbol $\hat{p}_{gu}$ is also decoded to retrieve its private part, $\hat{s}^{1}_{gu}$. Last, the $u$th user reencodes $\hat{s}^{0}_{gu}$ and $\hat{s}^{1}_{gu}$ and finally reconstruct its original message $\hat{s}_{gu}$. We implement in this work variations of this technique to mitigate SIC-related issues.

The above transmission strategy can be seen as an extension to dual-polarized massive MIMO systems of the single-polarized HRS technique from \cite{Dai16}, which relies on two layers of common messages. However, in contrast to the original scheme, we transmit one layer of common messages, i.e., only within each group. As shown in \cite{Dai16}, conveying a single layer of common messages delivers good performance in scenarios where groups of users are located in non-overlapping angular sectors, which is the case studied in this paper\footnote{When different groups overlap, linear precoding strategies become unable to cancel perfectly interference. In these scenarios, the idea of HRS in \cite{Dai16} of transmitting a second common message intended for all groups is beneficial and helps to mitigate uncanceled inter-group interference. The application of this more generalized HRS concept to our current dual-polarized RSMA schemes arises as a promising future direction.}.
Under these considerations, we focus on investigating the limitations of RSMA in dual-polarized systems, as well as developing strategies for improving the performance of the technique within each group. To this end, we exploit the polarization domain to propose and study three transmission approaches, as explained next.

\subsubsection{MIMO-RSMA with polarization multiplexing (MIMO-RSMA-PMUX)}
In this approach, instead of transmitting a superimposed stream, the private and common messages are transmitted through independent polarized data streams, i.e., each message is assigned to one polarization. Such a strategy will enable users to decode common and private messages without relying on SIC\footnote{It is noteworthy that MIMO-RSMA-PMUX can also be generalized to multiple layers of common streams if users employ more than one pair of dual-polarized antennas. Specifically, as proposed for single-polarized MIMO systems in \cite{Mishra22}, instead of transmitting a single data stream, the BS could transmit to each user a vector of common streams and a vector of private streams multiplexed in the polarization domain. As a result, users would need to apply some detection technique to each receive polarization, such as the minimum mean square error (MMSE) receiver, but SIC could still be completely removed. Nevertheless, specific features and performance advantages of such a possibility are still unclear and shall be addressed in future work.}. In addition to freeing the system from errors of imperfect SIC, users should be able to recover the common message without interference of the private messages, i.e., in ideal conditions. In practice, however, users will experience cross-polar interference, an issue that will be investigated in our analysis.

For simplicity and illustration purposes, in this work, the common message is assigned to the vertical polarization and the private messages are assigned to the horizontal polarization. Note that this polarization choice can be optimized (to maximize sum-rate, for instance), but such a possibility is left for future works where an in-depth investigation can be carried-out. With this strategy, the precoding vector for the common message can be written as $\mathbf{c}_{g} = [(\mathbf{c}^v_{g})^H, \mathbf{0}_{1, \frac{\bar{M}}{2}}]^H $, and for the private message as $\mathbf{p}_{gu} = [ \mathbf{0}_{1, \frac{\bar{M}}{2}}, (\mathbf{p}^h_{gu})^H]^H $. As a result, the signal transmitted by the BS in \eqref{eqsig1} can be rewritten as
\begin{align}\label{eq02}
    \mathbf{x} &= \sum_{m = 1}^{G}\mathbf{K}_{m}\left( \begin{bmatrix}
    \mathbf{c}^v_{m} \\
    \mathbf{0}_{\frac{\bar{M}}{2},1}
    \end{bmatrix}   \sqrt{\alpha_{m}} c_{m} \right. \nonumber\\
    &+ \left.
    \sum_{n = 1}^{U} \begin{bmatrix}
    \mathbf{0}_{\frac{\bar{M}}{2},1} \\
    \mathbf{p}^h_{mn} 
    \end{bmatrix} \sqrt{\beta_{mn}} p_{mn} \right)
    \in \mathbb{C}^{M \times 1},
\end{align}
where $\mathbf{c}^v_{m} \in \mathbb{C}^{\frac{\bar{M}}{2} \times 1}$ and $\mathbf{p}^h_{mn} \in \mathbb{C}^{\frac{\bar{M}}{2} \times 1}$ are the precoding vectors for the common and private messages corresponding to polarizations $v$ and $h$, respectively.

After the data streams in \eqref{eq02} have been transmitted though the channel in \eqref{eq01}, the $u$th user in the $g$th group will receive the following signal
\begin{align}\label{eq03}
    \mathbf{y}_{gu} &=  \mathbf{H}^H_{gu} \sum_{m = 1}^{G} \mathbf{K}_{m}
    \begin{bmatrix}
    \mathbf{c}^v_{m} \sqrt{\alpha_{m}} c_{m} \\
    \sum_{n = 1}^{U} \mathbf{p}^h_{mn} \sqrt{\beta_{mn}} p_{mn}
    \end{bmatrix} \nonumber\\
    &+
    \begin{bmatrix}
    n^v_{gu} \\ n^h_{gu} 
    \end{bmatrix} \in \mathbb{C}^{2\times 1},
\end{align}
where $n^i_{gu} \in \mathbb{C}$ is the additive white noise observed in polarization $i \in \{v, h\}$, which follows the complex Gaussian distribution with zero mean and variance $\sigma^2$.

\subsubsection{MIMO-RSMA with polarization diversity (MIMO-RSMA-PDIV)}
In the second approach, RSMA is employed per polarization. This means that replicas of the common and private messages will be transmitted in the two polarizations, allowing users to exploit diversity in the reception. Note, however, that differently from the first approach, the common message will still be detected with interference from the private messages, and SIC will still be required, possibly resulting in SIC errors. Nevertheless, as it will be demonstrated, polarization diversity will improve the overall performance of the system. Mathematically, the BS transmits
\begin{align}\label{eq04}
    \mathbf{x} &= \sum_{m = 1}^{G}\mathbf{K}_{m} \left( \begin{bmatrix}
    \mathbf{c}^v_{m} \\
    \mathbf{c}^h_{m} 
    \end{bmatrix} \sqrt{\alpha_{m}} c_{m}  \right. \nonumber\\
    &+ \left. 
    \sum_{n = 1}^{U}
    \begin{bmatrix}
    \mathbf{p}^v_{mn} \\
    \mathbf{p}^h_{mn} 
    \end{bmatrix} \sqrt{\beta_{mn}} p_{mn} \right)  \in \mathbb{C}^{M \times 1},
\end{align}
where $\mathbf{c}^i_{m} \in \mathbb{C}^{\frac{\bar{M}}{2} \times 1}$ and $\mathbf{p}^i_{mn} \in \mathbb{C}^{\frac{\bar{M}}{2} \times 1}$ are, respectively, the precoding vectors for the common and private messages transmitted in polarization $i \in \{v, h\}$.

As a result, the $u$th user in the $g$th group will receive the following signal
\begin{align}\label{eq05}
    \mathbf{y}_{gu} &=  \mathbf{H}^H_{gu} \sum_{m = 1}^{G} \mathbf{K}_{m}
    \begin{bmatrix}
    \mathbf{c}^v_{m} \sqrt{\alpha_{m}} c_{m} + \sum_{n = 1}^{U} \mathbf{p}^v_{mn} \sqrt{\beta_{mn}} p_{mn}\\
    \mathbf{c}^h_{m} \sqrt{\alpha_{m}} c_{m} + \sum_{n = 1}^{U} \mathbf{p}^h_{mn} \sqrt{\beta_{mn}} p_{mn}
    \end{bmatrix}  \nonumber\\
    &+
    \begin{bmatrix}
    n^v_{gu} \\ n^h_{gu} 
    \end{bmatrix} \in \mathbb{C}^{2\times 1}.
\end{align}

\subsubsection{MIMO-RSMA with SIC-aided polarization multiplexing (MIMO-RSMA-SPMUX)}
In the third approach, the conventional SIC-based RSMA technique is employed per polarization to multiplex two independent superimposed data streams. As a result, with the aid of SIC, each user will be able to recover two common and two private symbols, thus, leading to potential rate gains if compared to the previous strategies. Nevertheless, since this approach is based on the original RSMA protocol, users will experience the mentioned SIC-related problems. When the BS operates with the MIMO-RSMA-SPMUX scheme, the following signal is transmitted
\begin{align}\label{eq304}
    \mathbf{x} 
    & = \sum_{m = 1}^{G}\mathbf{K}_{m} \left( \begin{bmatrix}
    \mathbf{c}^v_{m} \sqrt{\alpha_{m}} c^v_{m}\\
    \mathbf{c}^h_{m} \sqrt{\alpha_{m}} c^h_{m}
    \end{bmatrix} \right. \nonumber\\
    &+ \left.
    \sum_{n = 1}^{U}
    \begin{bmatrix}
    \mathbf{p}^v_{mn} \sqrt{\beta_{mn}} p^v_{mn}\\
    \mathbf{p}^h_{mn} \sqrt{\beta_{mn}} p^h_{mn}
    \end{bmatrix}  \right)  \in \mathbb{C}^{M \times 1},
\end{align}
where $c^i_{m}$ and $p^i_{mn}$ represent, respectively, the common and private messages transmitted in polarization $i \in \{v, h\}$.

The $u$th user in the $g$th group will, then, receive
\begin{align}\label{eq305}
    \mathbf{y}_{gu} &= 
    \mathbf{H}^H_{gu} \sum_{m = 1}^{G} \mathbf{K}_{m}
    \begin{bmatrix}
    \mathbf{c}^v_{m} \sqrt{\alpha_{m}} c^v_{m} + \sum_{n = 1}^{U} \mathbf{p}^v_{mn} \sqrt{\beta_{mn}} p^v_{mn}\\
    \mathbf{c}^h_{m} \sqrt{\alpha_{m}} c^h_{m} + \sum_{n = 1}^{U} \mathbf{p}^h_{mn} \sqrt{\beta_{mn}} p^h_{mn}
    \end{bmatrix} \nonumber\\
    &+
    \begin{bmatrix}
    n^v_{gu} \\ n^h_{gu} 
    \end{bmatrix} \in \mathbb{C}^{2\times 1}.
\end{align}

\begin{figure}[t]
	\centering
	\includegraphics[width=1\linewidth]{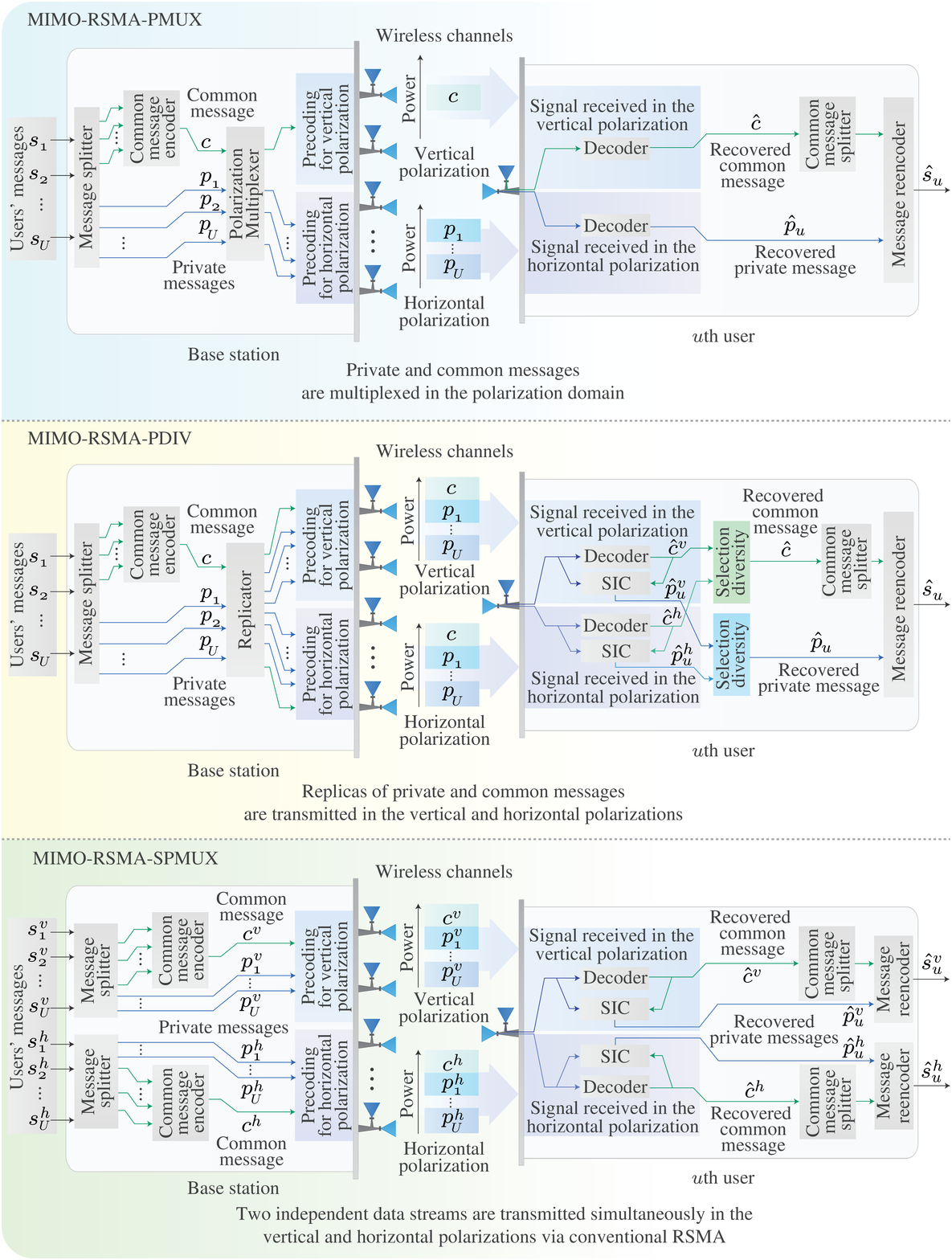}
	\caption{Transceivers for MIMO-RSMA-PMUX, MIMO-RSMA-PDIV, and MIMO-RSMA-SPMUX.}\label{f1.1}
\end{figure}
The three proposed transmission approaches are illustrated in Fig. \ref{f1.1}.

\section{Precoder Design}\label{secprec}
In order to employ RSMA to each group separately, the precoding matrix $\mathbf{K}_g \in \mathbb{C}^{M\times \bar{M}}$ should be designed to eliminate the inter-group interference. Thus, we are interested in achieving $\mathbf{H}_{gu}^H\mathbf{K}_{g'} = \mathbf{H}_{gu}^H \big(\mathbf{I}_{2} \otimes \mathbf{F}_{g'} \big) = \mathbf{0}_{2,\bar{M}}, \forall g \neq g'$. As can be noticed in \eqref{eq01}, such a goal can be accomplished by exploiting the null space spanned by the left eigenvector matrices of interfering groups, i.e., by defining $ \mathbf{U}^{*}_{g} = [\mathbf{U}_{1},\cdots,\mathbf{U}_{g-1}, \mathbf{U}_{g+1}, \cdots, \mathbf{U}_{G}] \in \mathbb{C}^{\frac{M}{2} \times \sum_{g'\neq g} \bar{r}_{g'}}$, $\mathbf{F}_{g} \in \mathbb{C}^{\frac{M}{2}\times\frac{\bar{M}}{2}}$ can be constructed from the orthonormal basis of $\mathrm{null}\{\mathbf{U}^{*}_{g}\}$, which can be obtained from the eigenvectors corresponding to the zero eigenvalues of $\mathbf{U}^{*}_{g}$. More specifically, let $\mathbf{E}^{0}_{g} \in \mathbb{C}^{\frac{M}{2}\times \frac{M}{2} - \sum_{g'\neq g} \bar{r}_{g'}}$ denote the matrix comprising the last $\frac{M}{2} - \sum_{g'\neq g} \bar{r}_{g'}$ left eigenvectors of $\mathbf{U}^{*}_{g}$ computed via SVD. Then, the desired precoding matrix can be given by $\mathbf{K}_{g} = \mathbf{I}_{2} \otimes \mathbf{F}_{g} = \mathbf{I}_{2} \otimes \big[\mathbf{E}^{0}_{g}\big]_{1:\frac{\bar{M}}{2}}$, in which, due to the dimension of $\mathbf{E}^{0}_{g}$, it is required that $\frac{\bar{M}}{2} \leq \frac{M}{2} - \sum_{g'\neq g} \bar{r}_{g'}$, $\frac{M}{2} > \sum_{g'\neq g} \bar{r}_{g'}$, and $\frac{\bar{M}}{2} \leq \bar{r}_{g} \leq r_{g}$. To satisfy these constraints, the parameter $\bar{r}_g$ is configured as $\mathrm{min}\{r_g, \lfloor (\frac{M}{2} - \frac{\bar{M}}{2})/(G  -1)  \rfloor \}$.

Given that the design of $\mathbf{K}_g$ only depends on the users' channel covariance matrices, which are slowly varying and can be estimated accurately via long-term measurements, it is reasonable to assume that $\mathbf{K}_{g}$ can cancel all inter-group interference. As a result, the signal for MIMO-RSMA-PMUX, in \eqref{eq03}, can be simplified as
\begin{align}
    \mathbf{y}_{gu} &=  \sqrt{\zeta_{gu}}
    \begin{bmatrix} (\mathbf{h}^{vv}_{gu})^H\mathbf{F}_{g} & \sqrt{\chi}(\mathbf{h}^{hv}_{gu})^H\mathbf{F}_{g} \\
    \sqrt{\chi}(\mathbf{h}^{vh}_{gu})^H\mathbf{F}_{g} & (\mathbf{h}^{hh}_{gu})^H\mathbf{F}_{g}\end{bmatrix} \nonumber\\
    &\times
     \begin{bmatrix}
    \mathbf{c}^v_{g} \sqrt{\alpha_{g}} c_{g} \\
    \sum_{n = 1}^{U} \mathbf{p}^h_{gn} \sqrt{\beta_{gn}} p_{gn}
    \end{bmatrix} +
    \begin{bmatrix}
    n^v_{gu} \\ n^h_{gu} 
    \end{bmatrix},
\end{align}
for MIMO-RSMA-PDIV, in \eqref{eq05}, as
\begin{align}
    \mathbf{y}_{gu} &=  \sqrt{\zeta_{gu}}
    \begin{bmatrix} (\mathbf{h}^{vv}_{gu})^H\mathbf{F}_{g} & \sqrt{\chi}(\mathbf{h}^{hv}_{gu})^H\mathbf{F}_{g} \\
    \sqrt{\chi}(\mathbf{h}^{vh}_{gu})^H\mathbf{F}_{g} & (\mathbf{h}^{hh}_{gu})^H\mathbf{F}_{g}\end{bmatrix} \nonumber\\
    &\times
    \begin{bmatrix}
    \mathbf{c}^v_{g} \sqrt{\alpha_{g}} c_{g} + \sum_{n = 1}^{U} \mathbf{p}^v_{gn} \sqrt{\beta_{gn}} p_{gn}\\
    \mathbf{c}^h_{g} \sqrt{\alpha_{g}} c_{g} + \sum_{n = 1}^{U} \mathbf{p}^h_{gn} \sqrt{\beta_{gn}} p_{gn}
    \end{bmatrix} +
    \begin{bmatrix}
    n^v_{gu} \\ n^h_{gu} 
    \end{bmatrix},
\end{align}
and for MIMO-RSMA-SPMUX, in \eqref{eq305}, as
\begin{align}
    \mathbf{y}_{gu} &=  \sqrt{\zeta_{gu}}
    \begin{bmatrix} (\mathbf{h}^{vv}_{gu})^H\mathbf{F}_{g} & \sqrt{\chi}(\mathbf{h}^{hv}_{gu})^H\mathbf{F}_{g} \\
    \sqrt{\chi}(\mathbf{h}^{vh}_{gu})^H\mathbf{F}_{g} & (\mathbf{h}^{hh}_{gu})^H\mathbf{F}_{g}\end{bmatrix} 
    \nonumber\\
    &\times
    \begin{bmatrix}
    \mathbf{c}^v_{g} \sqrt{\alpha_{g}} c^v_{g} + \sum_{n = 1}^{U} \mathbf{p}^v_{gn} \sqrt{\beta_{gn}} p^v_{gn}\\
    \mathbf{c}^h_{g} \sqrt{\alpha_{g}} c^h_{g} + \sum_{n = 1}^{U} \mathbf{p}^h_{gn} \sqrt{\beta_{gn}} p^h_{gn}
    \end{bmatrix} +
    \begin{bmatrix}
    n^v_{gu} \\ n^h_{gu} 
    \end{bmatrix}.
\end{align}
Now that the inter-group interference have been addressed, we can proceed and provide the details for the common and private precoders.
First let us concentrate on the design of the precoding vector for the private massages. The role of $\mathbf{p}^{i}_{gu} \in \mathbb{C}^{\frac{\bar{M}}{2} \times 1}$, $i \in \{v, h\}$, is to eliminate intra-group interference, i.e., to ensure that each private message reaches only its intended user in the assigned polarization. This means that both $\mathbf{p}^v_{gu}$ and $\mathbf{p}^h_{gu}$ for all transmission approaches can be designed in a similar fashion. More specifically, $\mathbf{p}^i_{gu}$ must be orthogonal to the subspace spanned by the effective channel\footnote{In our analysis, we assume that the BS can estimate perfectly the users' fast-varying channels. It is noteworthy that our main goal in this work is to investigate the fundamental performance impacts of imperfect SIC in RSMA systems and strategies to mitigate them. Nevertheless, an example with inaccurate CSI is also studied in our simulation results in Section V.} $(\mathbf{h}^{ii}_{gu'})^H\mathbf{F}_{g} \in \mathbb{C}^{1\times\frac{\bar{M}}{2}}$ of users $u' \neq u$, for $i\in\{v, h\}$. Therefore, the precoding vector for polarization $i\in \{v, h\}$ can be computed by
\begin{align}
    \mathbf{p}^i_{gu} = \mathrm{null}\{ \mathbf{F}^H_{g}[\mathbf{h}^{ii}_{g1}, &\cdots, \mathbf{h}^{ii}_{g(u-1)}, \nonumber\\
    &\mathbf{h}^{ii}_{g(u+1)}, \cdots, \mathbf{h}^{ii}_{gU}] \}  \in \mathbb{C}^{\frac{\bar{M}}{2} \times 1},
\end{align}
in which we must have $\bar{M}> U - 1$ to ensure the existence of a non-trivial null space.

On the other hand, $\mathbf{c}^{i}_{g} \in \mathbb{C}^{\frac{\bar{M}}{2} \times 1}$ should be designed to broadcast the common messages to all users. Different strategies and optimization procedures have been proposed in the literature, including matched filter precoding (MFP), weighted MFP, precoding for max-min fairness, among others \cite{Mao20, Mao18}. However, such sophisticated precoding designs can lead to an intractable analysis. As an alternative, for mathematical simplicity, we construct $\mathbf{c}^{i}_{g}$ in this work as a random precoder with independent and identically distributed entries following the standard complex Gaussian distribution, as in \cite{Dizdar21}.

With the above strategies, for both clustering the users and for constructing $\mathbf{F}_{g}$, the BS only needs the channel covariance matrices. Nevertheless, the BS also requires the knowledge of the effective channels $(\mathbf{h}^{ii}_{gu})^H\mathbf{F}_{g}$, for $i \in \{v,h\}$. As shown in \cite{ref6}, this effective CSI can be efficiently estimated at each coherence interval with a small feedback overhead. Moreover, it is noteworthy that the cross-polarized channels $\mathbf{h}^{ij}_{gu} \in \mathbb{C}^{\frac{M}{2} \times 1}$, for $i \neq j \in \{v,h\}$, are not required for our proposed transmission approaches, which further simplifies the CSI acquisition process.

\emph{Complexity remarks:} On the BS side, the complexity for computing the precoders for inter-group interference cancellation is the same for all dual-polarized MIMO-RSMA schemes, i.e., all proposed MA approaches employ the same outer precoding matrix $\mathbf{K}_g$. However, the complexity of the inner precoders differs among the schemes. Specifically, for each group, the MIMO-RSMA-PMUX scheme needs to compute one common precoder for the vertical polarization, $\mathbf{c}^v_{g}$, and $U$ private precoders for the horizontal polarization, $\mathbf{p}^h_{gu}$. On the other hand, both MIMO-RSMA-PDIV and MIMO-RSMA-SPMUX strategies require one common and $U$ private precoders for each polarization, i.e., $2(U + 1)$ inner precoders in total. Regarding the encoding process, in the MIMO-RSMA-PMUX and MIMO-RSMA-PDIV schemes, $U$ users' messages are split and encoded into $U + 1$ data symbols. In contrast, because MIMO-RSMA-SPMUX sends two parallel streams, $2U$ messages are encoded into $2(U + 1)$ data symbols for each group. Therefore, on the transmitter side, MIMO-RSMA-PMUX is the less complex scheme, whereas the MIMO-RSMA-SPMUX counterpart has the highest complexity. On the receiver side, MIMO-RSMA-PMUX also has the lowest computational complexity since it does not require SIC. The other two schemes, in their turn, both require one SIC layer per polarization. However, the MIMO-RSMA-PDIV scheme is slightly more complex than MIMO-RSMA-SPMUX since it implements selection diversity for recovering private and common symbols.

\section{Performance Analysis}

\subsection{Signal detection and SINR analysis for MIMO-RSMA-PMUX}
In this subsection, we determine the SINR for the first dual-polarized MIMO-RSMA transmission approach. To this end,
we assume that $\mathbf{p}^h_{gu}$ can successfully null out the inter-user interference observed in the corresponding polarization $h$. As a result, the $u$th user in the $g$th group detects the following signal
\begin{align}\label{detsgia1}
    \mathbf{y}_{gu} &=  
    \begin{bmatrix} \sqrt{\zeta_{gu}} (\mathbf{h}^{vv}_{gu})^H\mathbf{F}_{g}\mathbf{c}^v_{g} \sqrt{\alpha_{g}} c_{g} \\ \sqrt{\zeta_{gu}}
    (\mathbf{h}^{hh}_{gu})^H\mathbf{F}_{g} \mathbf{p}^h_{gu} \sqrt{\beta_{gu}} p_{gu} \end{bmatrix} \nonumber\\
    &+
    \begin{bmatrix} \sqrt{\zeta_{gu}\chi}(\mathbf{h}^{hv}_{gu})^H\mathbf{F}_{g} \sum_{n = 1}^{U} \mathbf{p}^h_{gn} \sqrt{\beta_{gn}} p_{gn} \\ \sqrt{\zeta_{gu}\chi}(\mathbf{h}^{vh}_{gu})^H\mathbf{F}_{g}\mathbf{c}^v_{g} \sqrt{\alpha_{g}} c_{g} \end{bmatrix}
    +
    \begin{bmatrix}
    n^v_{gu} \\ n^h_{gu} 
    \end{bmatrix}.
\end{align}

As can be observed in \eqref{detsgia1}, upon reception, users become able to recover the common and private messages directly from the assigned polarization, effortless without SIC. However, both messages are corrupted by cross-polar interference that $\mathbf{p}^h_{gu}$ is unable to cancel. Consequently, the SINR experienced by the $u$th in the $g$th group when detecting the common message can be given by
\begin{align}\label{sinr_a1_c}
    \gamma^{c}_{gu} &= \frac{ \zeta_{gu} |(\mathbf{h}^{vv}_{gu})^H\mathbf{F}_{g}\mathbf{c}^{v}_{g}|^2 \alpha_{g}}
    { \zeta_{gu}\chi \sum_{n=1}^{U}|(\mathbf{h}^{hv}_{gu})^H\mathbf{F}_{g} \mathbf{p}^{h}_{gn}|^2 \beta_{gn} + \sigma^2}.
\end{align}

Analogously, the SINR observed by the $u$th user when decoding its private message can be written as
\begin{align}\label{sinr_a1_p}
    \gamma^{p}_{gu} = \frac{ \zeta_{gu} |(\mathbf{h}^{hh}_{gu})^H\mathbf{F}_{g} \mathbf{p}^h_{gu}|^2 \beta_{gu} }
    {\zeta_{gu}\chi|(\mathbf{h}^{vh}_{gu})^H\mathbf{F}_{g}\mathbf{c}^v_{g}|^2 \alpha_{g}  + \sigma^2}.
\end{align}

\subsection{Signal detection and SINR analysis for MIMO-RSMA-PDIV}
Now, we determine the SINR for the second dual-polarized transmission strategy. Specifically, by also assuming that $\mathbf{p}^i_{gu}$ can successfully remove the inter-user interference in the corresponding polarization $i \in \{v, h\}$, the signal observed by the $u$th user in the $g$th group with MIMO-RSMA-PDIV can be expressed by 
\begin{align}\label{recsig01}
    &\mathbf{y}_{gu}
    = \begin{bmatrix} \sqrt{\zeta_{gu}}\left( (\mathbf{h}^{vv}_{gu})^H\mathbf{F}_{g} \mathbf{c}^v_{g} + \sqrt{\chi}(\mathbf{h}^{hv}_{gu})^H\mathbf{F}_{g} \mathbf{c}^h_{g} \right) \sqrt{\alpha_{g}} c_{g}   \\ \sqrt{\zeta_{gu}}\left( (\mathbf{h}^{hh}_{gu})^H\mathbf{F}_{g} \mathbf{c}^h_{g} + \sqrt{\chi}(\mathbf{h}^{vh}_{gu})^H\mathbf{F}_{g} \mathbf{c}^v_{g} \right)\sqrt{\alpha_{g}} c_{g}  \end{bmatrix} \nonumber\\
    &+
    \begin{bmatrix}  
    \sqrt{\zeta_{gu}}\left( (\mathbf{h}^{vv}_{gu})^H\mathbf{F}_{g} \mathbf{p}^v_{gu} 
    + \sqrt{\chi}(\mathbf{h}^{hv}_{gu})^H\mathbf{F}_{g} \mathbf{p}^h_{gu} 
    \right)\sqrt{\beta_{gu}} p_{gu}  \\  
    \sqrt{\zeta_{gu}} \left( (\mathbf{h}^{hh}_{gu})^H\mathbf{F}_{g} \mathbf{p}^h_{gu}
    + \sqrt{\chi}(\mathbf{h}^{vh}_{gu})^H\mathbf{F}_{g} \mathbf{p}^v_{gu}
    \right) \sqrt{\beta_{gu}} p_{gu} \end{bmatrix} \nonumber\\
    & + 
    \begin{bmatrix} \sqrt{\zeta_{gu}\chi}(\mathbf{h}^{hv}_{gu})^H\mathbf{F}_{g} \sum\nolimits_{n = 1, n\neq u }^{U} \mathbf{p}^h_{gn} \sqrt{\beta_{gn}} p_{gn} \\ \sqrt{\zeta_{gu}\chi}(\mathbf{h}^{vh}_{gu})^H\mathbf{F}_{g} \sum\nolimits_{n = 1, n\neq u}^{U} \mathbf{p}^v_{gn} \sqrt{\beta_{gn}} p_{gn} \end{bmatrix}
    +
    \begin{bmatrix}
    n^v_{gu} \\ n^h_{gu} 
    \end{bmatrix}.
\end{align}

In MIMO-RSMA-PDIV, we follow the original RSMA protocol and detect first the common message while treating the private message as interference. Note, however, that users are able to receive replicas of the common and private messages in the two polarizations. This means that, despite the existence of more interfering terms in \eqref{recsig01}, users can exploit diversity to improve their performance. Therefore, we employ a selection diversity strategy and recover the common message from the polarization with the strongest effective channel gain, which is denoted by $\varrho^{c,\star}_{gu} = \mathrm{max}\{\varrho^{c,v}_{gu}, \varrho^{c,h}_{gu}\}$, where $\varrho^{c,v}_{gu} = \zeta_{gu}(| (\mathbf{h}^{vv}_{gu})^H\mathbf{F}_{g} \mathbf{c}^v_{g}|^2 + \chi|(\mathbf{h}^{hv}_{gu})^H\mathbf{F}_{g} \mathbf{c}^h_{g} |^2) \alpha_{g}$ and $\varrho^{c,h}_{gu} = \zeta_{gu}(| (\mathbf{h}^{hh}_{gu})^H\mathbf{F}_{g} \mathbf{c}^h_{g}|^2 + \chi|(\mathbf{h}^{vh}_{gu})^H\mathbf{F}_{g} \mathbf{c}^v_{g} |^2 )\alpha_{g}$. As a result, the SINR for the common message experienced by the $u$th can be given by
\begin{align}\label{sinr_a2_c}
    \gamma^{c}_{gu} &= \frac{ \varrho^{c,\star}_{gu}}
    {\varpi^{c,\star}_{gu} + \sigma^2},
\end{align}
where $\varpi^{c,\star}_{gu}$ is the interference term defined in \eqref{sinr_interf_c_a2}, shown on the top of the next page.

\begin{figure*}[!t]
	% ensure that we have normalsize text
	\normalsize
	% Store the current equation number.
	%\setcounter{MYtempeqncnt}{\value{equation}}
	% Set the equation number to one less than the one
	% desired for the first equation here.
	% The value here will have to changed if equations
	% are added or removed prior to the place these
	% equations are referenced in the main text.
	\setcounter{equation}{17}
	\begin{align}
    \varpi^{c,\star}_{gu} &= 
    \begin{cases} 
        \zeta_{gu}(|(\mathbf{h}^{vv}_{gu})^H\mathbf{F}_{g} \mathbf{p}^v_{gu} |^2 \beta_{gu}
        + \chi\sum_{n = 1}^{U} |(\mathbf{h}^{hv}_{gu})^H\mathbf{F}_{g}  \mathbf{p}^h_{gn}|^2 \beta_{gn}), \quad \text{ if } \quad \varrho^{c,\star}_{gu} = \varrho^{c,v}_{gu}, \\
        \zeta_{gu}(|(\mathbf{h}^{hh}_{gu})^H\mathbf{F}_{g} \mathbf{p}^h_{gu} |^2 \beta_{gu}
        + \chi\sum_{n = 1}^{U} |(\mathbf{h}^{vh}_{gu})^H\mathbf{F}_{g}  \mathbf{p}^v_{gn}|^2 \beta_{gn}), \quad \text{ if } \quad \varrho^{c,\star}_{gu} = \varrho^{c,h}_{gu}.
    \end{cases}\label{sinr_interf_c_a2}\\
    \varpi^{p,\star}_{gu} &=
    \begin{cases} 
        \zeta_{gu}(\xi\varrho^{c,v}_{gu} + 
        \chi \sum_{n = 1, n\neq u }^{U} |(\mathbf{h}^{hv}_{gu})^H\mathbf{F}_{g} \mathbf{p}^h_{gn}|^2 \beta_{gn}), \quad \text{ if } \quad \varrho^\star_{gu} = \varrho^{p,v}_{gu}, \\
        \zeta_{gu}(\xi\varrho^{c,h}_{gu} + 
        \chi \sum_{n = 1, n\neq u }^{U} |(\mathbf{h}^{vh}_{gu})^H\mathbf{F}_{g} \mathbf{p}^v_{gn}|^2 \beta_{gn}), \quad \text{ if } \quad \varrho^\star_{gu} = \varrho^{p,h}_{gu}.
    \end{cases}\tag{20}\label{sinr_interf_p_a2}
\end{align}

	% Restore the current equation number.
	\setcounter{equation}{18}
	% The IEEE uses as a separator
	\hrulefill
	% The spacer can be tweaked to stop underfull vboxes.
\end{figure*}

Next, each user executes SIC in both polarizations to subtract the common message $c_g$ from the composite signal in \eqref{recsig01}. Then, similarly to the strategy employed to detect the common message, users recover the private messages from the polarization with the best effective channel gain. Ideally, SIC can remove the common message perfectly. However, as already discussed, several issues can corrupt the estimate of $c_g$ and lead to imperfect SIC. As in \cite{SenaISIC2020}, this residual SIC error is represented by a linear factor that models the uncancelled power of the common message, which, in practice, can be obtained by $\xi = \mathrm{E}\{ |c_g - \hat{c}_g|^2 \}$, with $\hat{c}_g$ denoting the estimate of ${c}_g$. Under this assumption, and denoting the best channel gain by $\varrho^{p,\star}_{gu} = \mathrm{max}\{\varrho^{p,v}_{gu}, \varrho^{p,h}_{gu}\}$, where $\varrho^{p,v}_{gu} = \zeta_{gu}(| (\mathbf{h}^{vv}_{gu})^H\mathbf{F}_{g} \mathbf{p}^v_{gu}|^2 
    + \chi |(\mathbf{h}^{hv}_{gu})^H\mathbf{F}_{g} \mathbf{p}^h_{gu} 
    |^2 )\beta_{gu}$ and $ \varrho^{p,h}_{gu} =  
    \zeta_{gu}(| (\mathbf{h}^{hh}_{gu})^H\mathbf{F}_{g} \mathbf{p}^h_{gu}|^2
    + \chi|(\mathbf{h}^{vh}_{gu})^H\mathbf{F}_{g} \mathbf{p}^v_{gu}
    |^2) \beta_{gu}$, the $u$th user in the $g$th group with will observe the following SINR when decoding its private message
\begin{align}\label{sinr_a2_p}
    \gamma^{p}_{gu} &= \frac{  \varrho^{p,\star}_{gu} }
    { \varpi^{p,\star}_{gu} + \sigma^2},
\end{align}
where $\varpi^{p,\star}_{gu}$ is the interference term defined in \eqref{sinr_interf_p_a2}, shown on the top of this page.

\subsection{Signal detection and SINR analysis for MIMO-RSMA-SPMUX}
The signal observed by the $u$th user in the $g$th group operating under the MIMO-RSMA-SPMUX scheme can be expressed by 
\setcounter{equation}{20}
\begin{align}\label{recsig301}
    &\mathbf{y}_{gu}
    = \begin{bmatrix} \sqrt{\zeta_{gu}\alpha_{g}}\left( (\mathbf{h}^{vv}_{gu})^H\mathbf{F}_{g} \mathbf{c}^v_{g} c^v_{g} + \sqrt{\chi}(\mathbf{h}^{hv}_{gu})^H\mathbf{F}_{g} \mathbf{c}^h_{g} c^h_{g} \right) \\ \sqrt{\zeta_{gu}\alpha_{g}}\left( (\mathbf{h}^{hh}_{gu})^H\mathbf{F}_{g} \mathbf{c}^h_{g} c^h_{g} + \sqrt{\chi}(\mathbf{h}^{vh}_{gu})^H\mathbf{F}_{g} \mathbf{c}^v_{g} c^v_{g} \right)   \end{bmatrix} \nonumber\\
    &+
    \begin{bmatrix}  
    \sqrt{\zeta_{gu}\beta_{gu}}\left( (\mathbf{h}^{vv}_{gu})^H\mathbf{F}_{g} \mathbf{p}^v_{gu} p^v_{gu}
    + \sqrt{\chi}(\mathbf{h}^{hv}_{gu})^H\mathbf{F}_{g} \mathbf{p}^h_{gu} p^h_{gu}
    \right)  \\  
    \sqrt{\zeta_{gu}\beta_{gu}} \left( (\mathbf{h}^{hh}_{gu})^H\mathbf{F}_{g} \mathbf{p}^h_{gu} p^h_{gu}
    + \sqrt{\chi}(\mathbf{h}^{vh}_{gu})^H\mathbf{F}_{g} \mathbf{p}^v_{gu}  p^v_{gu}
    \right) \end{bmatrix} \nonumber\\
    & + 
    \begin{bmatrix} \sqrt{\zeta_{gu}\chi}(\mathbf{h}^{hv}_{gu})^H\mathbf{F}_{g} \sum\nolimits_{n = 1, n\neq u }^{U} \mathbf{p}^h_{gn} \sqrt{\beta_{gn}} p^h_{gn} \\ \sqrt{\zeta_{gu}\chi}(\mathbf{h}^{vh}_{gu})^H\mathbf{F}_{g} \sum\nolimits_{n = 1, n\neq u}^{U} \mathbf{p}^v_{gn} \sqrt{\beta_{gn}} p^v_{gn} \end{bmatrix}
    +
    \begin{bmatrix}
    n^v_{gu} \\ n^h_{gu} 
    \end{bmatrix}.
\end{align}

Then, users follow the original RSMA protocol and recover the desired common messages directly from the corresponding polarization. Specifically, in this first stage of the decoding process, the $u$th user detects the common message $c^i_{g}$ in polarization $i \in \{v,h\}$ under interference from the private message $p^i_{gu}$, and also from $p^j_{gu}$ and $c^j_{g}$, for $i\neq j \in \{v,h\}$, that have leaked from the orthogonal polarization as a result of depolarization phenomena. Thus, the SINR for the common message detected by the $u$th user in polarization $i\in\{v, h\}$ can be expressed by
\begin{align}\label{sinr_a3_c}
    \gamma^{c,i}_{gu} &= \frac{ \varrho^{c,i}_{gu}}
    {\varpi^{c,i}_{gu} + \sigma^2},
\end{align}
where $\varrho^{c,i}_{gu} = \zeta_{gu}| (\mathbf{h}^{ii}_{gu})^H\mathbf{F}_{g} \mathbf{c}^i_{g}|^2\alpha_{g}$, and $\varpi^{c,i}_{gu}$ is the term modeling the interference observed in polarization $i$, defined by
\begin{align}\label{sinr_interf_c_a3}
    \varpi^{c,i}_{gu} &= \zeta_{gu}\Bigg(|(\mathbf{h}^{ii}_{gu})^H\mathbf{F}_{g} \mathbf{p}^i_{gu} |^2 \beta_{gu}
        + \chi|(\mathbf{h}^{ji}_{gu})^H\mathbf{F}_{g} \mathbf{c}^j_{g} |^2 \alpha_{g} \nonumber\\
        &\left. + \chi\sum_{n = 1}^{U} |(\mathbf{h}^{ji}_{gu})^H\mathbf{F}_{g}  \mathbf{p}^j_{gn}|^2 \beta_{gn}\right),
\end{align}
for $i\neq j\in\{v,h\}$. After executing SIC and possibly introducing decoding errors, the $u$th user in the $g$th group recovers its private message in polarization $i\in \{v,h\}$ with the following SINR
\begin{align}\label{sinr_a3_p}
    \gamma^{p,i}_{gu} &= \frac{  \varrho^{p,i}_{gu} }
    { \varpi^{p,i}_{gu} + \sigma^2},
\end{align}
where $\varrho^{p,i}_{gu} = \zeta_{gu}| (\mathbf{h}^{ii}_{gu})^H\mathbf{F}_{g} \mathbf{p}^i_{gu}|^2 \beta_{gu}$, and $\varpi^{p,i}_{gu}$ models the interference in polarization $i$, given by
\begin{align}\label{sinr_interf_p_a3}
    \varpi^{p, i}_{gu} &= \zeta_{gu}\Bigg(\xi\varrho^{c,i}_{gu} + \chi|(\mathbf{h}^{ji}_{gu})^H\mathbf{F}_{g} \mathbf{c}^j_{g} |^2 \alpha_{g}  \nonumber\\
    &\left. + 
        \chi \sum_{n = 1}^{U} |(\mathbf{h}^{ji}_{gu})^H\mathbf{F}_{g} \mathbf{p}^j_{gn}|^2 \beta_{gn}\right),
\end{align}
for $i\neq j\in\{v,h\}$, where $\xi$ is the SIC error factor.

\subsection{Statistical characterization of channel gains for MIMO-RSMA-PMUX}\label{statc_a1}
We perform in this subsection an statistical characterization on the channel gains of the SINR expressions in \eqref{sinr_a1_c} and \eqref{sinr_a1_p}. First, let us concentrate on the SINR for the common message. Let the gain in the numerator of \eqref{sinr_a1_c} be denoted by $\varsigma^c_{gu} = \zeta_{gu} |(\mathbf{h}^{vv}_{gu})^H\mathbf{F}_{g}\mathbf{c}^{v}_{g}|^2 \alpha_{g}$ and the interference term by $\omega^c_{gu} = \zeta_{gu}\chi \sum_{n=1}^{U}|(\mathbf{h}^{hv}_{gu})^H\mathbf{F}_{g} \mathbf{p}^{h}_{gn}|^2 \beta_{gn}$. Then, the squared norm in $\varsigma^c_{gu}$ can be expanded as $|(\mathbf{h}^{vv}_{gu})^H\mathbf{F}_{g}\mathbf{c}^{v}_{g}|^2 = [(\mathbf{h}^{vv}_{gu})^H\mathbf{F}_{g}\mathbf{c}^{v}_{g}]^H(\mathbf{h}^{vv}_{gu})^H\mathbf{F}_{g}\mathbf{c}^{v}_{g} = (\mathbf{c}^{v}_{g})^H\mathbf{F}^H_{g}\mathbf{h}^{vv}_{gu}(\mathbf{h}^{vv}_{gu})^H\mathbf{F}_{g}\mathbf{c}^{v}_{g}
= \mathrm{tr}\{\mathbf{F}^H_{g}\mathbf{h}^{vv}_{gu}(\mathbf{h}^{vv}_{gu})^H\mathbf{F}_{g}\mathbf{c}^{v}_{g}(\mathbf{c}^{v}_{g})^H\}$. Given that $\mathbf{c}^{v}_{g}$ is an isotropic unit vector independent of $(\mathbf{h}^{vv}_{gu})^H\mathbf{F}_{g}$ and that $\mathrm{E}\{\mathbf{c}^{v}_{g}(\mathbf{c}^{v}_{g})^H\} = \frac{1}{\bar{M}}\mathbf{I}_{\bar{M}}$, we have that $\mathrm{E}\{|(\mathbf{h}^{vv}_{gu})^H\mathbf{F}_{g}\mathbf{c}^{v}_{g}|^2 \} = \frac{1}{\bar{M}}\mathrm{tr}\{\mathbf{F}^H_{g}\mathbf{h}^{vv}_{gu}(\mathbf{h}^{vv}_{gu})^H\mathbf{F}_{g}\}$.
Since $\mathbf{g}^{vv}_{gu}$ in \eqref{eq01} is a complex Gaussian distributed vector and $\mathbf{F}_{g}$ is a semi-unitary matrix (with orthonormal columns), the vector $\mathbf{F}^H_{g}\mathbf{h}^{vv}_{gu}$ is still complex Gaussian distributed.
Consequently, the gain $\varsigma^c_{gu} = \zeta_{gu} |(\mathbf{h}^{vv}_{gu})^H\mathbf{F}_{g}\mathbf{c}^{v}_{g}|^2 \alpha_{g}$ follows a Gamma distribution with shape parameter $1$ and rate parameter $\phi/\zeta_{gu}\alpha_{g}$, with $\phi = \bar{M}/\mathrm{tr}\{ \mathbf{F}^H_{g}\mathbf{R}_{g}\mathbf{F}_{g}\}$, whose probability distribution function (PDF) can be found in \cite[eq. (15.1.1)]{Krishnamoorthy2006}.

Next, let us focus on the interference term $\omega^c_{gu}$. Note that $\omega^c_{gu}$ consists of a sum of $U$ correlated random variables (RVs). As a result, determining the exact distribution of $\omega^c_{gu}$ becomes challenging. For overcoming this issue, we approximate the PDF of $\omega^c_{gu}$ by assuming that the sum terms are independent. Then, as it was performed for $\varsigma^c_{gu}$, we expand the squared norm in $\omega^c_{gu}$ as $\mathrm{tr}\{\mathbf{F}^H_{g}\mathbf{h}_{gu}^{hv}(\mathbf{h}_{gu}^{hv})^{H}\mathbf{F}_{g}\mathbf{p}^{h}_{gn}(\mathbf{p}^{h}_{gn})^H\}$. Since $\mathbf{p}^{h}_{gn}$ is uniformly distributed with the Haar measure $O(\bar{M})$, we have that $\mathrm{E}\{\mathbf{p}^{h}_{gn}(\mathbf{p}^{h}_{gn})^H\} = \frac{1}{\bar{M}}\mathbf{I}_{\bar{M}}$. As a result, the distribution of each term of the summation in $\omega^c_{gu}$ follows a Gamma distribution with shape parameter $1$ and rate parameter ${\phi}/{\zeta_{gu} \chi\beta_{gn}}$, for $n=1,\cdots,U$. This means that $\omega^c_{gu}$ consists of a sum of Gamma RVs with shape 1, which is known to follow a Hypoexponential distribution \cite{Amari97}. However, since we employ in this work a uniform power allocation among the private messages\footnote{As stated in \cite{Dizdar21}, uniform power allocation for multi-user MIMO is widely employed in the literature and used in practical 4G and 5G systems.}, we have that $\beta_{g1} = \beta_{g2} = \cdots = \beta_{gU}$. This implies that the terms of the sum in $\omega^c_{gu}$ have equal rate parameters. Consequently, the PDF of $\omega^c_{gu}$ can be approximated by a Gamma PDF with shape parameter $U$ and rate parameter ${\phi}/{\zeta_{gu} \chi\beta_{gu}}$. 

We characterize now the gains in the SINR for the private messages in \eqref{sinr_a1_p}. Let $\varsigma^{p}_{gu} = \zeta_{gu} |(\mathbf{h}^{hh}_{gu})^H\mathbf{F}_{g} \mathbf{p}^h_{gu}|^2 \beta_{gu}$ and $\omega^{p}_{gu} = \zeta_{gu}\chi|(\mathbf{h}^{vh}_{gu})^H\mathbf{F}_{g}\mathbf{c}^v_{g}|^2 \alpha_{g}$. As can be noticed, $\varsigma^{p}_{gu}$ and $\omega^{p}_{gu}$ have a similar form of $\varsigma^c_{gu}$, which implies that such gains can be characterized similarly as we have characterized $\varsigma^c_{gu}$. Therefore, the full details for this statistical characterization are omitted. In short, $\varsigma^{p}_{gu}$ and $\omega^{p}_{gu}$ follow Gamma distributions with shape parameters $1$ and rate parameters given by $\phi/\zeta_{gu}\beta_{gu}$ and $\phi/\zeta_{gu} \chi\alpha_{g}$, respectively.

\begin{figure*}[!t]
	% ensure that we have normalsize text
	\normalsize
	% Store the current equation number.
	%\setcounter{MYtempeqncnt}{\value{equation}}
	% Set the equation number to one less than the one
	% desired for the first equation here.
	% The value here will have to changed if equations
	% are added or removed prior to the place these
	% equations are referenced in the main text.
	\setcounter{equation}{28}
	\begin{align}\label{pdf_interf_p_a2}
    f_{\varpi^{p, \star}_{gu}}(y) 
    & = \frac{\phi \left(1 - \frac{ \beta_{gu}}{\xi\alpha_{g}}\right)^{-U+1} e^{-y \frac{\phi}{\zeta_{gu} \xi\chi \alpha_{g}}} \gamma\left(U - 1, y \phi \left[ (\zeta_{gu} \chi \beta_{gu})^{-1} - (\zeta_{gu} \xi\chi \alpha_{g})^{-1} \right]\right)}{\xi(\chi - 1)\Gamma(U-1) \zeta_{gu} \alpha_{g}}
    \nonumber\\
    &- 
    \frac{\phi \left(1 - \frac{\chi \beta_{gu} }{\xi \alpha_{g}}\right)^{-U+1} e^{-y \frac{\phi}{\zeta_{gu} \xi \alpha_{g}}} \gamma\left(U - 1, y \phi \left[ (\zeta_{gu} \chi \beta_{gu})^{-1} - (\zeta_{gu} \xi \alpha_{g})^{-1} \right]\right)}{\xi(\chi - 1)\Gamma(U-1) \zeta_{gu} \alpha_{g}}.
\end{align}

	% Restore the current equation number.
	\setcounter{equation}{25}
	% The IEEE uses as a separator
	\hrulefill
	% The spacer can be tweaked to stop underfull vboxes.
\end{figure*}

\subsection{Statistical characterization of channel gains for MIMO-RSMA-PDIV}\label{statc_a2}
As in the previous subsection, the gains in \eqref{sinr_a2_c} and \eqref{sinr_a2_p} are also correlated. Therefore, for mathematical tractability, we also make the assumption of independence between the gains in this subsection. Given this assumption, let us start by characterizing $\varrho^{c,\star}_{gu} = \mathrm{max}\{\varrho^{p,v}_{gu}, \varrho^{p,h}_{gu}\}$ in \eqref{sinr_a2_c}. It can be easily verified that $\varrho^{c,v}_{gu}$ and $\varrho^{c,h}_{gu}$ are identically distributed, each one consisting of a sum of two Gamma distributed RVs, with each term having shape parameters $1$ and rate parameters $ \phi / \zeta_{gu} \alpha_{g} $ and ${\phi}/{\zeta_{gu}\chi\alpha_{g}}$. Therefore, the gains $\varrho^{c,i}_{gu}$, for $i\in \{v,h\}$, follow a two-parameter Hypoexponential distribution with PDF given by \cite{Amari97}
\begin{align}\label{two_p_hypo_pdf}
    f_{\varrho^c_{gu}}(x) &= \frac{\phi}{(1 - \chi)\zeta_{gu} \alpha_{g}}\left(e^{-x \frac{ \phi}{\zeta_{gu} \alpha_{g}}} - e^{-x \frac{\phi}{\zeta_{gu} \chi \alpha_{g}}} \right).
\end{align}

Given that $\varrho^{c,v}_{gu}$ and $\varrho^{c,h}_{gu}$ are identically distributed, the PDF of $\varrho^{c,\star}_{gu}$ can be obtained by $f_{\varrho^{c, \star}_{gu}}(x) = 2 f_{\varrho^c_{gu}}(x) F_{\varrho^c_{gu}}(x)$, where $F_{\varrho^c_{gu}}(x)$ denotes the cumulative distribution function (CDF) of $\varrho^{c,i}_{gu}$, which can be easily obtained by integrating \eqref{two_p_hypo_pdf}. Thus, the desired PDF is obtained as
\begin{align}\label{pdf_gain_c_a2}
    f_{\varrho^{c, \star}_{gu}}(x) &= \frac{2 \phi }{\zeta_{gu} \alpha_{g}} \left[\frac{e^{-x  \frac{\phi}{\zeta_{gu} \alpha_{g}}}}{1 - \chi} - \frac{e^{-x \frac{ \phi }{\zeta_{gu} \chi \alpha_{g}}}}{1 - \chi} - \frac{\chi e^{-x \frac{2 \phi}{\zeta_{gu} \chi \alpha_{g}}}}{(1 - \chi)^2} \right.\nonumber\\
    &\left.  - \frac{e^{-x \frac{2\phi}{\zeta_{gu} \alpha_{g}} }}{(1 - \chi)^2}  + \frac{(1 + \chi)e^{-x  \frac{\phi(1 + \chi)}{ \zeta_{gu} \chi \alpha_{g}}}}{(1 - \chi)^2} \right].
\end{align}

Regarding the interference term $\varpi^{c,\star}_{gu}$, it can be verified that the underlying RVs in \eqref{sinr_interf_c_a2} are identically distributed. This means that the distribution of $\varpi^{c,\star}_{gu}$ will be equal to the distribution of the cases in \eqref{sinr_interf_c_a2}, regardless of $\varrho^{c, \star}_{gu} = \varrho^{c,v}_{gu}$ or $ \varrho^{c, \star}_{gu} = \varrho^{c,h}_{gu}$. More specifically, each case consists of a sum of two RVs, with the first one following a Gamma distribution with shape parameter 1 and rate parameter $\phi / \zeta_{gu}\beta_{gu}$, and the second one following a Gamma distribution with shape parameter $U$ and rate parameter ${\phi}/{\zeta_{gu}\chi\beta_{gu}}$. Consequently, the PDF of $\varpi^{c,\star}_{gu}$ can be approximated by the convolution between the PDFs of these two sum terms, yielding
\begin{align}\label{pdf_interf_c_a2}
    f_{\varpi^{c, \star}_{gu}}(y) & = \frac{\phi}{(1-\chi)^U \Gamma(U) \zeta_{gu} \beta_{gu}} e^{-y \frac{\phi }{ \zeta_{gu} \beta_{gu}}} \nonumber\\
    &\times \gamma\left(U, y \phi\left[(\zeta_{gu} \chi \beta_{gu})^{-1} - (\zeta_{gu} \beta_{gu})^{-1} \right]\right).
\end{align}

Similar steps can be carried out for characterizing the gains in the SINR of private messages in \eqref{sinr_a2_p}. Specifically, $\varrho^{p,v}_{gu}$ and $\varrho^{p,h}_{gu}$ can be also approximated as two independent and identically distributed two-parameter Hypoexponential RVs with rate parameters ${\phi}/{\zeta_{gu}\beta_{gu}}$ and ${\phi}/{\zeta_{gu}\chi\beta_{gu}}$. As a result, the PDF of $\varrho^{p,\star}_{gu} = \mathrm{max}\{\varrho^{p,v}_{gu}, \varrho^{p,h}_{gu}\}$ can be obtained directly from \eqref{pdf_gain_c_a2} by replacing $\alpha_{g}$ by $\beta_{gu}$. Therefore, the PDF expression for $\varrho^{p,\star}_{gu}$ will not be reproduced due to space constraints.

The interference term $\varpi^{p,\star}_{gu}$ in \eqref{sinr_interf_p_a2}, in its turn, consists of a sum of SIC interference (with residual interference of the common message) and cross-polar interference with the private messages of other users. Specifically, the interference term containing the residual SIC error can be characterized by a Hypoexponential distribution with rate parameters ${\phi}/{\zeta_{gu}\xi\alpha_{g}}$ and ${\phi}/{\zeta_{gu}\xi\chi\alpha_{g}}$, whereas the cross-polar interference term follows a Gamma distribution with shape parameter $U-1$ and rate parameter ${\phi}/{\zeta_{gu}\chi\beta_{gu}}$. The PDF for $\varpi^{p,\star}_{gu}$ can, thus, be approximated by the convolution between the PDFs of the two interference terms, which results in the expression in \eqref{pdf_interf_p_a2}, shown on the top of this page.

\subsection{Statistical characterization of channel gains for MIMO-RSMA-SPMUX}
The effective channel gains for the third approach are now characterized. First, as can be observed, the gain $\varrho^{c,i}_{gu} = \zeta_{gu} |(\mathbf{h}^{ii}_{gu})^H\mathbf{F}_{g}\mathbf{c}^{i}_{g}|^2 \alpha_{g}$ in the SINR for the common message in \eqref{sinr_a3_c} is identical to the gain in \eqref{sinr_a1_c}. Therefore, from subsection \ref{statc_a1}, we have that $\varrho^{c,i}_{gu}$ follows a Gamma distribution with shape parameter $1$ and rate parameter $\phi/\zeta_{gu}\alpha_{g}$, with $\phi = \bar{M}/\mathrm{tr}\{ \mathbf{F}^H_{g}\mathbf{R}_{g}\mathbf{F}_{g}\}$ and PDF given in \cite[eq. (15.1.1)]{Krishnamoorthy2006}.
In its turn, the interference term $\varpi^{c,i}_{gu}$ consists of a sum of three correlated RVs. From the characterization provided in the previous subsection, it is easy to verify that the sum of the two first terms in \eqref{sinr_interf_c_a3} follows a Hypoexponential distribution with parameters $ \phi / \zeta_{gu} \beta_{gu} $ and ${\phi}/{\zeta_{gu}\chi\alpha_{g}}$ and thus, has a PDF similar to \eqref{two_p_hypo_pdf}. On the other hand, the third interference term follows a Gamma distribution with shape parameter $U$ and rate parameter ${\phi}/{\zeta_{gu}\chi\beta_{gu}}$. As a result, the PDF for $\varpi^{p,i}_{gu}$ can be approximated by the convolution between the Hypoexponential and Gamma PDFs, similarly as in \eqref{pdf_interf_p_a2}.

The statistical characterization for the SINR expression of the private messages in \eqref{sinr_a3_p} can be performed following similar steps as above. In a few words, the gain $\varrho^{p,i}_{gu} = \zeta_{gu} |(\mathbf{h}^{ii}_{gu})^H\mathbf{F}_{g}\mathbf{p}^{i}_{gu}|^2 \beta_{gu}$ follows a Gamma distribution with shape parameter $1$ and rate parameter $\phi/\zeta_{gu}\beta_{gu}$, whereas the interference term $\varpi^{p,i}_{gu}$ can be characterized by the convolution between the PDFs of a Hypoexponential distribution, with parameters ${\phi}/{\zeta_{gu}\xi\alpha_{g}}$ and ${\phi}/{\zeta_{gu}\chi\alpha_{g}}$, and a Gamma distribution with shape parameter $U$ and rate parameter ${\phi}/{\zeta_{gu}\chi\beta_{gu}}$.

\begin{figure*}[!t]
	% ensure that we have normalsize text
	\normalsize
	% Store the current equation number.
	%\setcounter{MYtempeqncnt}{\value{equation}}
	% Set the equation number to one less than the one
	% desired for the first equation here.
	% The value here will have to changed if equations
	% are added or removed prior to the place these
	% equations are referenced in the main text.
	\setcounter{equation}{31}
	\begin{align}
    P^c_{gu} &= \begin{cases}
    1 - 2 &\hspace{-3mm} \left[\frac{\alpha_{g}^{U+1} (\alpha_{g} + \chi\beta_{gu}\tau_g^c)^{-U}}{(1 - \chi)(\alpha_{g} + \beta_{gu}\tau_g^c)} e^{-\frac{\phi\tau_g^c}{\rho\zeta_{gu}\alpha_{g}}} %\right.\nonumber\\
    %&
    - \frac{\chi^2\alpha_{g}^{U+1} (\alpha_{g} + \beta_{gu}\tau_g^c)^{-U}}{(1 - \chi)(\chi\alpha_{g} + \beta_{gu}\tau_g^c) } e^{-\frac{\phi\tau_g^c}{\rho\zeta_{gu}\chi\alpha_{g}}} \right. \\
    &- \frac{\chi^3\alpha_{g}^{U+1} (\alpha_{g} + 2\beta_{gu}\tau_g^c)^{-U} }{2(1-\chi)^2(\chi\alpha_{g} + 2\beta_{gu}\tau_g^c) } e^{-\frac{2\phi\tau_g^c}{\rho\zeta_{gu}\chi\alpha_{g}}} %\nonumber\\
    %&
    - \frac{\alpha_{g}^{U+1} (\alpha_{g} + 2\chi\beta_{gu}\tau_g^c)^{-U}}{2(1-\chi)^2(\alpha_{g} + 2\beta_{gu}\tau_g^c) }  e^{-\frac{2\phi\tau_g^c}{\rho\zeta_{gu}\alpha_{g}}} \\
     &+ \left. 
    \frac{\chi^2\alpha_{g}^{U+1} (\alpha_{g} + (1 + \chi)\beta_{gu}\tau_g^c)^{-U}}{(1-\chi)^2(\chi\alpha_{g} + (1 + \chi)\beta_{gu}\tau_g^c) }  e^{-\frac{\phi\tau_g^c\left(1 + \chi\right)}{\rho\zeta_{gu}\chi\alpha_{g}}} \right], \hfill \text{if} \quad \chi \neq 0,\\[3mm]
    1 - 2 &\hspace{-3mm}\left[ \frac{ \alpha_{g} }{\alpha_{g} + \beta_{gu}\tau_g^c} e^{-\frac{\phi\tau_g^c}{\rho\zeta_{gu}\alpha_{g}}} 
    %\right. \nonumber\\
    %&
    %\left.
    - \frac{\alpha_{g}}{2(\alpha_{g} + 2\beta_{gu}\tau_g^c) }  e^{-\frac{2\phi\tau_g^c}{\rho\zeta_{gu}\alpha_{g}}} \right], \hfill \text{otherwise.}
    \end{cases}\label{out_comm_a2}\\
    P^p_{gu} & = \begin{cases} 1  -  2 &\hspace{-3mm} \left[ \frac{ \beta_{gu}^2 (1 - \chi)^{-1} \left(1 + \chi \tau_{gu}^p \right)^{-U+1}}{\chi (\xi \alpha_{g} \tau_{gu}^p + \beta_{gu}) (\xi \alpha_{g} \tau_{gu}^p + \chi^{-1} \beta_{gu} )} e^{-\frac{\phi \tau_{gu}^p}{\rho\zeta_{gu}\beta_{gu} } } 
    - \frac{\chi^2 \beta_{gu}^2 (1 - \chi)^{-1} \left(1 + \tau_{gu}^p \right)^{-U+1}}{(\xi \alpha_{g} \tau_{gu}^p + \beta_{gu})(\xi \alpha_{g} \tau_{gu}^p + \chi\beta_{gu})}  e^{- \frac{\phi \tau_{gu}^p}{\rho\zeta_{gu}\chi\beta_{gu} } }
    \right.\\
    % & 
    % \nonumber\\
    &- \frac{\chi^3 \beta_{gu}^2 (1-\chi)^{-2} \left(1 + 2 \tau_{gu}^p \right)^{-U+1}}{2 ({2\xi \alpha_{g} \tau_{gu}^p + \beta_{gu}}) ({2\xi \alpha_{g} \tau_{gu}^p + \chi \beta_{gu}})} e^{- \frac{2\phi \tau_{gu}^p}{\rho\zeta_{gu}\chi\beta_{gu} } } 
    %\nonumber\\
    %&
    - \frac{\beta_{gu}^2 \chi (1-\chi)^{-2} }{2 ({2\xi  \alpha_{g} \tau_{gu}^p + \beta_{gu}})}
    \frac{\left(1 + 2\chi \tau_{gu}^p \right)^{-U+1}}{({2\xi \alpha_{g} \tau_{gu}^p + \chi^{-1}\beta_{gu} })} e^{-\frac{2\phi \tau_{gu}^p}{\rho\zeta_{gu}\beta_{gu} } }  
    \\
    &\left. + \frac{\chi^2 \beta_{gu}^2 (1-\chi)^{-2}}{(\xi \alpha_{g} \tau_{gu}^p(1 + \chi) + \beta_{gu})} \frac{\left(1 + (1 + \chi) \tau_{gu}^p \right)^{-U+1}}{(\xi \alpha_{g} \tau_{gu}^p(1 + \chi) + \chi\beta_{gu})}  e^{- \frac{\phi \tau_{gu}^p(1 + \chi)}{\rho\zeta_{gu}\chi\beta_{gu} } } \right], \hspace{5mm} \text{if} \quad \chi \neq 0,\\[3mm]
    1 - 2 &\hspace{-3mm} \left[ \frac{\beta_{gu}}{\xi \alpha_{g} \tau_{gu}^p + \beta_{gu}} e^{-\frac{\phi\tau_{gu}^p }{\rho\zeta_{gu}\beta_{gu} } } - \frac{ \beta_{gu} }{2({2\xi  \alpha_{g} \tau_{gu}^p + \beta_{gu}})}  e^{-\frac{2\phi\tau_{gu}^p }{\rho\zeta_{gu}\beta_{gu} } }  \right], \hspace{9mm} \text{otherwise.}
    \end{cases}\label{out_private_a2}
\end{align}

	% Restore the current equation number.
	%\setcounter{equation}{33}
	% The IEEE uses as a separator
	\hrulefill
	% The spacer can be tweaked to stop underfull vboxes.
\end{figure*}

\subsection{Outage probability}

In RSMA,  users need to successfully decode both common and private messages for them to be able to reconstruct the intended original information. Therefore, an outage event will occur either if the rate of the common message or the rate of the private message drops below the corresponding target data rate. Mathematically, the outage probability for the $u$th user in the $g$th group can be obtained by $P^{\text{out}}_{gu} = P^c_{gu} \cup P^p_{gu} = P^c_{gu} + P^p_{gu} - P^c_{gu} P^p_{gu}$,
where $P^c_{gu} = \mathrm{Pr}\{\log_2(1 + \gamma^{c}_{gu}) < R^{c}_{g}\}$ and $P^p_{gu} = \mathrm{Pr}\{\log_2(1 + \gamma^{p}_{gu}) < R^{p}_{gu}\}$, with $R^{c}_{g}$ and $R^{p}_{gu}$ representing, respectively, the target data rates for the common and private messages, which are measured in bits per channel use (bpcu).
Tight closed-form approximations for the outage probabilities of the proposed dual-polarized MIMO-RSMA schemes are derived in the following propositions.

\paragraph*{Proposition I} When the BS employs MIMO-RSMA-PMUX, the outage probability for the common message experienced by the $u$th user in the $g$th group can be approximated by
\setcounter{equation}{29}
\begin{align}\label{out_comm_a1}
    P^c_{gu} &= \text{\small $1 - \left(\frac{\alpha_{g}}{\alpha_{g} + \chi\beta_{gu}\tau_g^c } \right)^U  e^{-\frac{\phi\tau_g^c}{\rho\zeta_{gu}\alpha_{g}}}$},
\end{align}
where $\tau^{c}_{g} = 2^{R^{c}_{g}} - 1$ and $\rho = 1/\sigma^2$ is the signal-to-noise ratio (SNR).

\textit{Proof:} Please, see Appendix \ref{ap1}.  \hfill $\blacksquare$

\paragraph*{Proposition II} When the BS employs MIMO-RSMA-PMUX, the outage probability for the private message experienced by the $u$th user in the $g$th group can be approximated by
\begin{align}\label{out_private_a1}
    P^p_{gu} & = \text{\small $1 - \frac{\beta_{gu}}{ \chi\alpha_{g}\tau_{gu}^p + \beta_{gu}} e^{-\frac{\phi \tau_{gu}^p}{\rho\zeta_{gu}\beta_{gu}}}$},
\end{align}
where $\tau_{gu}^p = 2^{R_{gu}^{p}} - 1$. 

\textit{Proof:} Please, see Appendix \ref{ap2}.  \hfill $\blacksquare$

From the expressions provided in propositions I and II, it can be noticed that the outage probability for this strategy is degraded only when cross-polar interference exists. That is, the outage probabilities will saturate when $\rho \rightarrow \infty$ only if $\chi>0$. On the other hand, when $\chi = 0$, the outage probabilities should converge to zero as $\rho \rightarrow \infty$. As it will be shown in Section \ref{si_res_sec}, this behavior brings advantages to MIMO-RSMA-PMUX.

\paragraph*{Proposition III} When the BS employs MIMO-RSMA-PDIV, the outage probability for the common message experienced by the $u$th user in the $g$th group can be approximated by \eqref{out_comm_a2}, shown on the top of this page.

\textit{Proof:} Please, see Appendix \ref{ap4}.  \hfill $\blacksquare$

\paragraph*{Proposition IV} When the BS employs MIMO-RSMA-PDIV, the outage probability for the private message experienced by the $u$th user in the $g$th group can be approximated by \eqref{out_private_a2}, shown on the top of this page.

\textit{Proof:} Please, see Appendix \ref{ap5}.  \hfill $\blacksquare$

\begin{figure*}[!t]
	% ensure that we have normalsize text
	\normalsize
	% Store the current equation number.
	%\setcounter{MYtempeqncnt}{\value{equation}}
	% Set the equation number to one less than the one
	% desired for the first equation here.
	% The value here will have to changed if equations
	% are added or removed prior to the place these
	% equations are referenced in the main text.
	\setcounter{equation}{36}
	
	\begin{align}\label{erg_comm_a1}
    C^c_g &=  \sum_{u=1}^{U} \frac{(-1)^{U^2 - 1}}{\mathrm{ln}(2)}  \left(\frac{\alpha_{g}}{\chi\beta_{gu} - \alpha_{g}} \right)^{U^2} e^{\frac{\phi \sum_{l=1}^{U} \frac{1}{\zeta_{gl}} }{\rho \alpha_{g}} } \left[ \mathrm{Ei}\left(- \frac{\phi \sum_{l=1}^{U} \frac{1}{\zeta_{gl}} }{\rho \alpha_{g}}  \right) 
    %\nonumber\\
    %&
    - \bm{e}_{U^2-1}\left( \frac{\phi(\chi \beta_{gu} - \alpha_{g}) \sum_{l=1}^{U} \frac{1}{\zeta_{gl}}}{ \chi \rho \beta_{gu} \alpha_{g}}\right) \right.\nonumber\\
    & \times e^{- \frac{\phi(\chi \beta_{gu} - \alpha_{g}) \sum_{l=1}^{U} \frac{1}{\zeta_{gl}}}{ \chi \rho \beta_{gu} \alpha_{g}} } \mathrm{Ei}\left(- \frac{\phi \sum_{l=1}^{U} \frac{1}{\zeta_{gl}}}{\rho \chi \beta_{gu} }  \right) %\nonumber\\
    %& 
    + e^{-\frac{\phi \sum_{l=1}^{U} \frac{1}{\zeta_{gl}} }{\rho \alpha_{g}} } \sum_{m = 1}^{U^2 - 1} \frac{1}{m!} \left( - \frac{\chi \beta_{gu} - \alpha_{g} }{\alpha_{g} }\right)^{m} \nonumber\\
    & \left. \times \sum_{k = 0}^{m-1} (m-k-1)! \left(- \frac{\phi \sum_{l=1}^{U} \frac{1}{\zeta_{gl}}}{\rho \chi \beta_{gu} }  \right)^K \right].
\end{align}

	% Restore the current equation number.
	%\setcounter{equation}{42}
	% The IEEE uses as a separator
	\hrulefill
	% The spacer can be tweaked to stop underfull vboxes.
\end{figure*}

\paragraph*{Proposition V} When the BS employs MIMO-RSMA-SPMUX, the outage probability for the common message experienced by the $u$th user in the $g$th group can be approximated by
\setcounter{equation}{33}
\begin{align}\label{out_comm_a3}
    P^c_{gu} &= 1 - \frac{ \alpha_{g} }{(\chi \tau_g^c + 1)(\alpha_{g} + \beta_{gu}\tau_g^c)} \nonumber\\ 
    & \times \left(1 + \frac{\chi \beta_{gu}\tau_g^c}{\alpha_{g}}\right)^{-U} e^{-\frac{\phi\tau_g^c}{\rho\zeta_{gu}\alpha_{g}}}.
\end{align}

\textit{Proof:} The proof is similar as in Appendix C and is omitted due to space constraints.  \hfill $\blacksquare$

\paragraph*{Proposition VI} When the BS employs MIMO-RSMA-SPMUX, the outage probability for the private message experienced by the $u$th user in the $g$th group can be approximated by
\begin{align}\label{out_private_a3}
    P^p_{gu} & =  1 - \frac{\beta_{gu}^2}{(\xi \alpha_{g} \tau_{gu}^p + \beta_{gu})(\chi \alpha_{g} \tau_{gu}^p + \beta_{gu})}\nonumber\\
    & \times
    \left(1 + \chi \tau_{gu}^p\right)^{-U}e^{-\frac{\phi\tau_{gu}^p }{\rho\zeta_{gu}\beta_{gu} } }.
\end{align}

\textit{Proof:} The proof is similar as in Appendix D and is omitted due to space constraints.  \hfill $\blacksquare$

In contrast to the outage probabilities achieved for MIMO-RSMA-PMUX, the expressions in propositions IV and VI are also affected by the residual SIC error factor $\xi$. This implies that MIMO-RSMA-PDIV and MIMO-RSMA-SPMUX should experience more interference than MIMO-RSMA-PMUX. Moreover, as a result of the SIC protocol, \eqref{out_comm_a2} and \eqref{out_comm_a3} show that the outage probability for the common messages of MIMO-RSMA-PDIV and MIMO-RSMA-SPMUX, respectively, saturates when $\rho\rightarrow\infty$ even if $\chi=0$. Nevertheless, as a result of polarization diversity in MIMO-RSMA-PDIV, the exponential terms in \eqref{out_comm_a2} are multiplied by a factor of $2$, which suggests a faster decay in the outage probabilities. As it will be demonstrated in Section \ref{si_res_sec}, this benefit provides performance gains for the MIMO-RSMA-PDIV.

\subsection{Ergodic sum-rates}
In this subsection, we derive analytical expressions for the ergodic rates of the proposed RSMA schemes. In particular, for ensuring a successful decoding at all users, the ergodic rate for the common message is computed by averaging the minimum of the common message's instantaneous rates achieved by the users \cite{Mao18}. More specifically, the ergodic sum-rate for the $g$th group can be computed by
\begin{align}\label{ergsm_g_rsma}
C_g &=  C^p_{g} + C^c_{g},
\end{align}
where $C^p_{g} = \mathrm{E}\left\{  \sum_{u=1}^{U} \log_2\left(1 + \gamma^p_{gu}\right)\right\}$ and $C^c_{g} = \mathrm{E}\left\{ \sum_{u=1}^{U} \min_{l\in \{1, \cdots, U\} }\left\{ \log_2\left(1 + \gamma^c_{gl}\right) \right\} \right\}$ are the ergodic sum-rates for the private and common messages, respectively. We derive in the following propositions tight approximations for the ergodic sum-rates of the proposed approaches.

\paragraph*{Proposition VII} When the BS employs MIMO-RSMA-PMUX, the ergodic sum-rate for the common message experienced in the $g$th group can be approximated by \eqref{erg_comm_a1}, shown on the top of this page.

\textit{Proof:} Please, see Appendix \ref{ap6}.  \hfill $\blacksquare$

\paragraph*{Proposition VIII} When the BS employs MIMO-RSMA-PMUX, the ergodic sum-rate for the private message experienced in the $g$th group can be approximated by
\setcounter{equation}{37}
\begin{align}\label{erg_pri_a1}
    C^p_g &=  \sum_{u=1}^{U} \frac{ \beta_{gu}}{ \mathrm{ln}(2)(\beta_{gu} - \chi \alpha_{g})} \left[ e^{\frac{\phi}{\rho \zeta_{gu}\chi\alpha_{g}}} \mathrm{Ei}\left(- \frac{ \phi}{\rho \zeta_{gu}\chi\alpha_{g}} \right) \right. \nonumber\\
    &\left. - e^{\frac{\phi}{\rho \zeta_{gu}\beta_{gu} }} \mathrm{Ei}\left(- \frac{ \phi}{\rho \zeta_{gu} \beta_{gu} } \right)  \right].
\end{align}

\textit{Proof:} Please, see Appendix \ref{ap7}.  \hfill $\blacksquare$

\paragraph*{Proposition IX}  When the BS employs MIMO-RSMA-PDIV, the ergodic sum-rate for the private message experienced in the $g$th group can be approximated by \eqref{erg_pri_a2}, where $\Theta(\cdot)$ is defined in \eqref{erg_pri_a2_sub}, with $\Phi(\cdot)$ given in \eqref{sol_e_int}, and $\bar{\Phi}(\cdot)$ in \eqref{toprate_01}, shown on the top of the next page.

\textit{Proof:} Please, see Appendix \ref{ap8}.  \hfill $\blacksquare$

\begin{figure*}[!t]
	% ensure that we have normalsize text
	\normalsize
	% Store the current equation number.
	%\setcounter{MYtempeqncnt}{\value{equation}}
	% Set the equation number to one less than the one
	% desired for the first equation here.
	% The value here will have to changed if equations
	% are added or removed prior to the place these
	% equations are referenced in the main text.

\setcounter{equation}{38}
\begin{align}
    C^p_{g}  & = \sum_{u=1}^{U}  \left[ \frac{ 2\beta_{gu}}{ \chi (1 - \chi)^2} \Theta \left(\xi \alpha_{g}, \beta_{gu}, \chi, \frac{1}{\chi}, \frac{ \phi}{\rho \zeta_{gu} \beta_{gu} }, U \right) 
    %\nonumber\\
    %&
    - \frac{ 2\chi^2 \beta_{gu} }{ (1 - \chi)^2 } \Theta \left( \xi \alpha_{g}, \beta_{gu}, 1, \chi, \frac{ \phi}{\chi \rho \zeta_{gu} \beta_{gu} }, U \right)  \right.  \nonumber\\
    & - \frac{\chi^3 \beta_{gu} }{ (1-\chi)^3 } \Theta \left(2\xi \alpha_{g}, \beta_{gu}, 2, \chi,  \frac{ 2\phi}{\chi\rho \zeta_{gu} \beta_{gu} }, U \right)  
    %\nonumber\\
    %
    %&
    - \frac{ \beta_{gu} }{  \chi (1-\chi)^3} \Theta \left(2\xi \alpha_{g}, \beta_{gu}, 2\chi, \frac{1}{\chi}, \frac{ 2 \phi}{\rho \zeta_{gu} \beta_{gu} }, U \right)  \nonumber\\
    &\left.  + \frac{2 \chi^2 \beta_{gu} }{ (1-\chi)^3 }  \Theta \left((1 + \chi)\xi \alpha_{g}, \beta_{gu}, 1 + \chi, \chi, \frac{ (1 + \chi)\phi}{\chi \rho \zeta_{gu} \beta_{gu} }, U \right) \right].\label{erg_pri_a2}\\
    \Theta(a,b,c,d,h,l) &= \frac{1}{\mathrm{ln}(2)} \left[-\frac{e^{\frac{b h}{a}} \mathrm{Ei}\left( - \frac{b h}{a} \right)}{a - b}
     % \nonumber\\
    %&
    + \frac{e^{\frac{b h d }{a}} \mathrm{Ei}\left( - \frac{b h d }{a} \right)}{a - d b} + \frac{(1 - d) b e^{h} \mathrm{Ei}\left( - h \right)}{(a - b) (a - d b)} \right.
    %\nonumber\\
    %& 
    - \frac{(l-1) e^{\frac{b h}{a}} \Phi\left(\frac{h}{c}, \frac{ h (b c - a) }{ac}, 1, l-2 \right)}{a - b}  \nonumber\\
    & +  \frac{(l-1) e^{\frac{b h d}{a}} \Phi\left(\frac{h}{c}, \frac{ h (b c d - a) }{ac}, 1, l-2 \right)}{a - d b} 
     + \frac{(l-1) (1 - d) b e^{h} \bar{\Phi}\left(\frac{h}{c}, \frac{h(c - 1)}{c}, 1, l-2 \right)}{(a - b) (a - db)} \Bigg].\label{erg_pri_a2_sub}\\
    \Phi(\mu ,\nu ,\tau, n) &= -\frac{1}{n+1} \left[\frac{1}{\tau^{n+1}} + (-1)^n \left( \frac{\mu}{\nu} \right)^{n+1} \right]  \mathrm{Ei}(-\mu\tau - \nu) + \frac{(-1)^n}{n+1} \left( \frac{\mu}{\nu} \right)^{n+1} \bm{e}_{n}(\nu)e^{-\nu} \nonumber\\
    & \times \mathrm{Ei}(-\mu\tau) - \frac{(-1)^n}{n+1} \left( \frac{\mu}{\nu} \right)^{n+1} e^{-\mu\tau - \nu} \sum_{m=1}^{n} \frac{1}{m!} \left(-\frac{\nu}{\mu\tau}\right)^m \sum_{k=0}^{m-1} (m-k-1)! (-\mu\tau)^k.\label{sol_e_int}\\
    \bar{\Phi}(\mu , \nu, \tau, n) &= \begin{cases} \frac{1}{n + 1} \Big( - \left[ \frac{1}{\tau^{n+1}} + \frac{(-1)^n \mu^{n+1}}{(n+1)!}  \right] \mathrm{Ei}(-\mu \tau) - \frac{e^{-\mu\tau}}{(n+1)!\tau^{n+1} }  \sum_{m = 0}^{n} (n - m)! (-\mu\tau)^{m} \Big), & \text{if} \quad \nu=0, \\
        \Phi(\mu,\nu,n), & \text{otherwise.}
    \end{cases}\label{toprate_01}\\
    C^c_g &= \sum_{u=1}^{U}\frac{2^{U-1} \alpha_{g}^{U+1}}{U \mathrm{ln}(2)} \left[\frac{\Psi\left(\alpha_{g}, \chi\beta_{gu}, \beta_{gu}, \frac{\phi \sum_{l=1}^{U} \zeta_{gi} }{U\rho\alpha_{g}},U \right)}{1 - \chi} 
    %\nonumber\\
    %&
    - \frac{\chi \Psi\left(\alpha_{g}, \beta_{gu}, \frac{\beta_{gu}}{\chi}, \frac{\phi \sum_{l=1}^{U} \zeta_{gi} }{U\rho\chi\alpha_{g}},U \right)}{(1 - \chi)} \right.
    \nonumber\\
    &
    - \frac{\chi^2 \Psi\left(\alpha_{g}, 2\beta_{gu}, \frac{2\beta_{gu}}{\chi}, \frac{2 \phi \sum_{l=1}^{U} \zeta_{gi} }{U\rho\chi\alpha_{g}},U \right) }{2(1-\chi)^2} 
    %\nonumber\\
    %&
    - \frac{\Psi\left(\alpha_{g}, 2\chi\beta_{gu}, 2\beta_{gu}, \frac{2 \phi \sum_{l=1}^{U} \zeta_{gi} }{U\rho\alpha_{g}},U \right)}{2(1-\chi)^2}
    \nonumber\\
    &  
    + \left. \frac{\chi \Psi\left(\alpha_{g}, (1+\chi)\beta_{gu}, \frac{(1+\chi)\beta_{gu}}{\chi}, \frac{\phi (1+\chi)\sum_{l=1}^{U} \zeta_{gi} }{U\rho\chi\alpha_{g}},U \right) }{(1-\chi)^2 }  \right],\label{erg_comm_a2}\\
    \Psi (a,b,c,h,l) &= \frac{a^{-l}}{a-c} \left[ e^{\frac{ah}{c}}\mathrm{Ei} \left( -\frac{ah}{c} \right) - e^{h} \mathrm{Ei}(-h)  \right] %\nonumber\\
    %&
    +  \frac{l e^{h} }{a-c} \Phi\left(\frac{h}{b}, \frac{ h (b - a) }{b}, a, l-1 \right) \nonumber\\
    &
    - \frac{l e^{\frac{ah}{c}}}{a-c} \Phi\left(\frac{h}{b}, \frac{ ah (b - c) }{bc}, a, l-1 \right).\label{erg_comm_a2_sub}
\end{align}

	% Restore the current equation number.
	%\setcounter{equation}{42}
	% The IEEE uses as a separator
	\hrulefill
	% The spacer can be tweaked to stop underfull vboxes.
\end{figure*}

\paragraph*{Proposition X} When the BS employs MIMO-RSMA-PDIV, the ergodic sum-rate for the common message experienced in the $g$th group can be approximated by \eqref{erg_comm_a2}, where $\Psi (\cdot)$ is defined in \eqref{erg_comm_a2_sub}, and $\Phi(\cdot)$ is given in \eqref{sol_e_int}, shown on the top of the next page.

\textit{Proof:} The proof is similar as in Appendix \ref{ap8} and is omitted due to space constraints. \hfill $\blacksquare$

Expressions for the ergodic sum-rates of the proposed MIMO-RSMA-SPMUX scheme can be also derived following similar steps of propositions VII to X. However, due to space limitations, we are unable to provide such an analysis. Still, the ergodic sum-rates achieved with the MIMO-RSMA-SPMUX are investigated through insightful simulation examples in the following section.

\section{Numerical and Simulation Results}\label{si_res_sec}
The numerical and simulation examples presented in this section validate the theoretical analysis developed in the previous sections. Simulation results also confirm that the proposed dual-polarized MIMO-RSMA strategies can achieve impressive improvements in terms of outage probability, outage sum-rate, and ergodic sum-rate over conventional MA schemes. Specifically, we compare the performance of our proposals with single-polarized and dual-polarized baseline systems. Among the single-polarized schemes, we implement the legacy MIMO-OMA, which employs TDMA, the single-polarized MIMO-SDMA, the single-polarized MIMO-RSMA, and the single-polarized MIMO-NOMA. For the dual-polarized systems, we consider the dual-polarized MIMO-NOMA with polarization diversity from \cite{ni3} and dual-polarized MIMO-SDMA schemes, implementing both polarization diversity and multiplexing. In single-polarized systems, a single data message is encoded and transmitted by the BS. This implies that the single-polarized RSMA scheme sends one common and one private stream to the users. In the case of dual-polarized MIMO-SDMA and MIMO-NOMA schemes with polarization diversity, the BS transmits two replicas of a single data stream. On the other hand, the dual-polarized MIMO-SDMA with polarization multiplexing conveys two independent streams simultaneously, as the MIMO-RSMA-SPMUX does. Other specific simulation parameters are configured as follows.

\begin{figure*}[t]
	\centering
	\includegraphics[width=.75\linewidth]{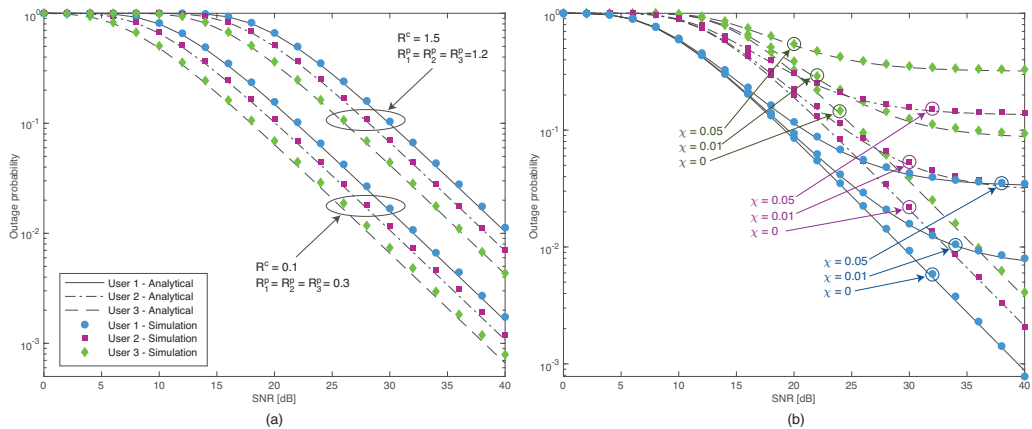}
	\caption{Analytical and simulated outage probabilities for MIMO-RSMA-PMUX: (a) for different target rates with $\chi = 0$, and (b) for different values of $\chi$ with $R^c = 0.5$~bpcu, $R^p_{1} = 0.1$~bpcu, $R^p_{2} = 0.5$~bpcu, and $R^p_{3} = 1.2$~bpcu.}\label{f2}
\end{figure*}

\begin{figure*}[t]
	\centering
	\includegraphics[width=.75\linewidth]{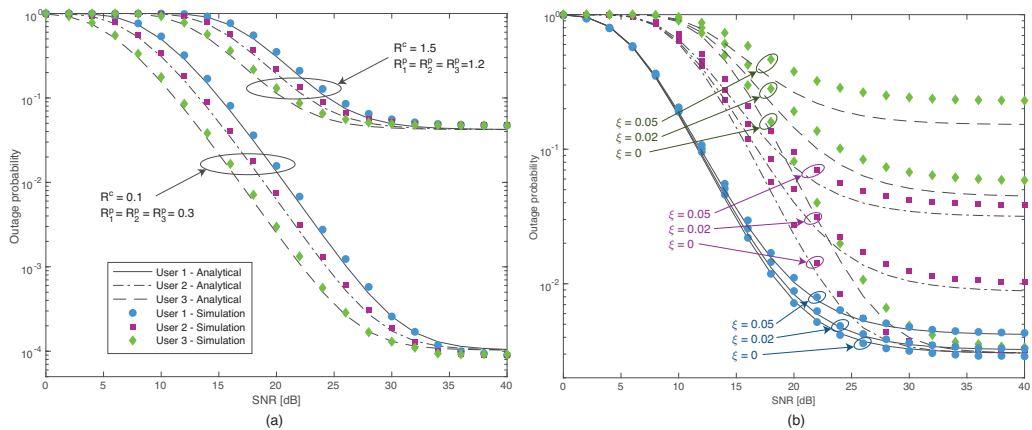}
	\caption{Analytical and simulated outage probabilities for MIMO-RSMA-PDIV: (a) for different target rates with $\chi = 0$ and $\xi = 0$, and (b) for different values of $\xi$ with $\chi = 0.001$ and $R^c = 0.5$~bpcu, $R^p_{1} = 0.1$~bpcu, $R^p_{2} = 0.5$~bpcu, and $R^p_{3} = 1.2$~bpcu.}\label{f3}
\end{figure*}

The dual-polarized schemes employ at the BS a uniform linear array with $\frac{M}{2} = 50$ co-located pairs of dual-polarized antennas, which implies a total of $M=100$ transmit antennas, whereas users employ one pair of dual-polarized antennas. The single-polarized systems, in their turn, implement the same number of antennas at the BS, and users employ one single-polarized receive antenna. The covariance matrices for all systems are generated using the one-ring scattering model \cite{ref6,ref1}.
To this end, we assume that users are distributed among $G = 4$ spatial groups surrounding the BS, with the $g$th group positioned at the azimuth angle given by $\theta_{g} = 30^\circ + (g-1) 160^\circ$ and a distance of $170$~m from the BS to its center. Furthermore, each group has a radius of $30$~m, which results in an angular spread of $\approx 10^\circ$. In particular, the presented results are based on the first group, which is located at the azimuth angle of $30^\circ$. Within each group there are $U=3$ users, in which users $1$, $2$, and $3$ are located, respectively, at $d_1 = 200$~m, $d_2 = 170$~m, and $d_3 = 140$~m from the BS. Under this setup, the large-scale fading coefficient for the $u$th user is modeled by $\zeta_{u} = \delta d_{u}^{-\eta}$, where $\delta$ is a BS array gain parameter that is adjusted based on the desired users' performance, and $\eta$ is the path-loss exponent. These parameters are configured as $\delta  = 4 \times 10^4$ and $\eta = 2.5$. Moreover, we set $\bar{M} = 6$ and adopt a fixed power allocation\footnote{More sophisticated adaptive power allocation strategies shall be considered in future work.}, where, for the RSMA schemes, we adjust $\alpha = 0.7$ and $\beta_u = (1 - \alpha)/U = 0.1$, for $u=1,\cdots, 3$. For the NOMA schemes, the power coefficients of users $1, 2$, and $3$ are set to $5/8, 2/8$, and $1/8$, respectively, whereas for OMA, the BS transmits at each time-slot using its full power, i.e., the users' coefficients are set to $1$.

Figs. \ref{f2}(a) and \ref{f2}(b) compare the analytical and simulated users' outage probabilities achieved with the dual-polarized MIMO-RSMA-PMUX scheme. Fig. \ref{f2}(a) shows outage probability curves for different sets of target rates considering a scenario with negligible cross-polar interference, i.e., for $\chi = 0$. It can be seen in this scenario that, independently of the specified target rates, the approximate analytical curves generated with \eqref{out_comm_a1} and \eqref{out_private_a1} can follow the simulated ones with high accuracy. In Fig. \ref{f2}(b), the performance of MIMO-RSMA-PMUX is tested for different levels of cross-polar interference, and an accurate agreement between simulated and analytical results can also be observed. In this figure, the outage probabilities of all users degrade and become limited in high SNR as $\chi$ increases. This behavior is expected and explained by the fact that private precoders become unable to cancel the inter-user interference perfectly when $\chi > 0$.

\begin{figure*}[t]
	\centering
	\includegraphics[width=.75\linewidth]{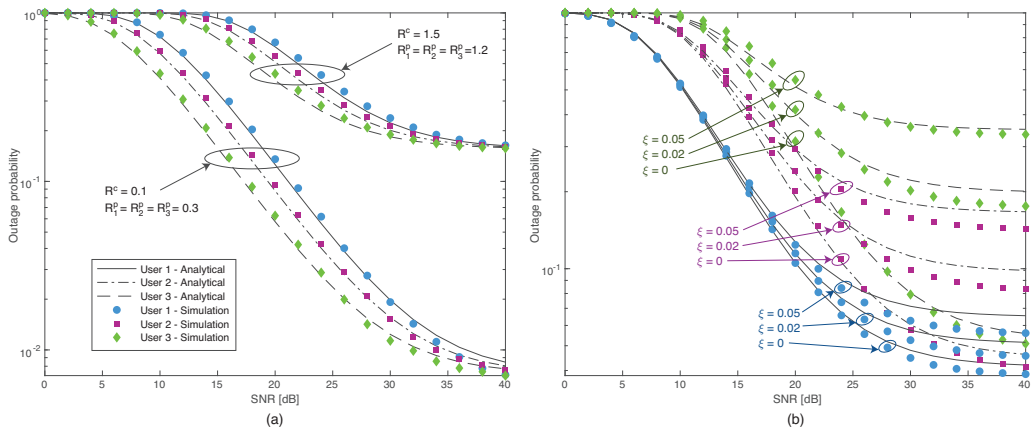}
	\caption{Analytical and simulated outage probabilities for MIMO-RSMA-SPMUX: (a) for different target rates with $\chi = 0$ and $\xi = 0$, and (b) for different values of $\xi$ with $\chi = 0.001$ and $R^c = 0.5$~bpcu, $R^p_{1} = 0.1$~bpcu, $R^p_{2} = 0.5$~bpcu, and $R^p_{3} = 1.2$~bpcu.}\label{rev_f1}
\end{figure*}

\begin{figure*}[t]
	\centering
	\includegraphics[width=.75\linewidth]{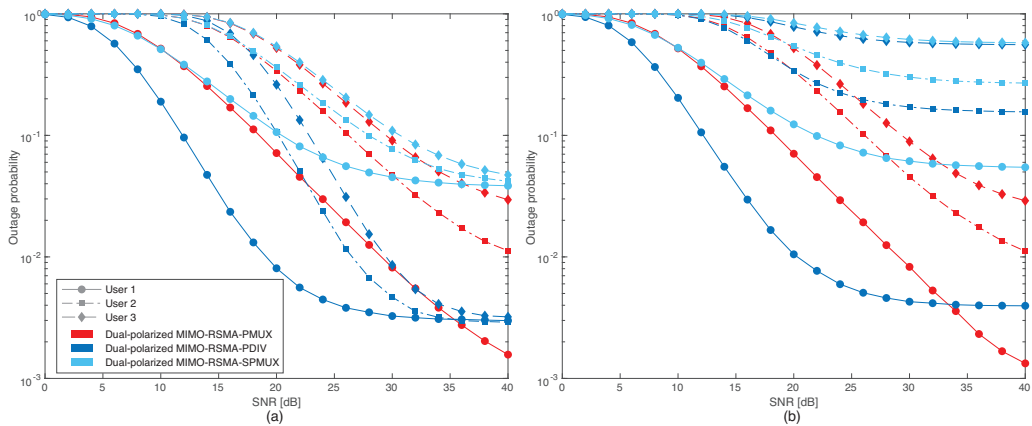}
	\caption{Simulated outage probabilities for the proposed dual-polarized RSMA schemes: (a) for a SIC error factor of $\xi = 0$, and (b) for $\xi = 0.05$ ($R^c = 0.5, R^p_{1} = 0.1, R^p_{2} = 1, R^p_{3} = 2$~bpcu, $\chi = 0.001$).}\label{rev_f2}
\end{figure*}

In Fig. \ref{f3}(a), we test the accuracy of our analytical results for MIMO-RSMA-PDIV considering different sets of target rates in a scenario with no cross-polar interference and no SIC errors. As can be seen, for this ideal case, the approximate analytical outage probabilities (generated with \eqref{out_comm_a2} and \eqref{out_private_a2}) can match almost perfectly the simulated ones in both sets of target rates. On the other hand, as can be seen in Fig. \ref{f2}(b), the accuracy of the analytical outage probability is impacted when cross-polar interference and imperfect SIC are considered. However, the analytical results still provide good approximations and perfectly capture the behavior and slope of the simulated outage probability curves, which corroborates our analysis.

\begin{figure*}[t]
	\centering
	\includegraphics[width=.75\linewidth]{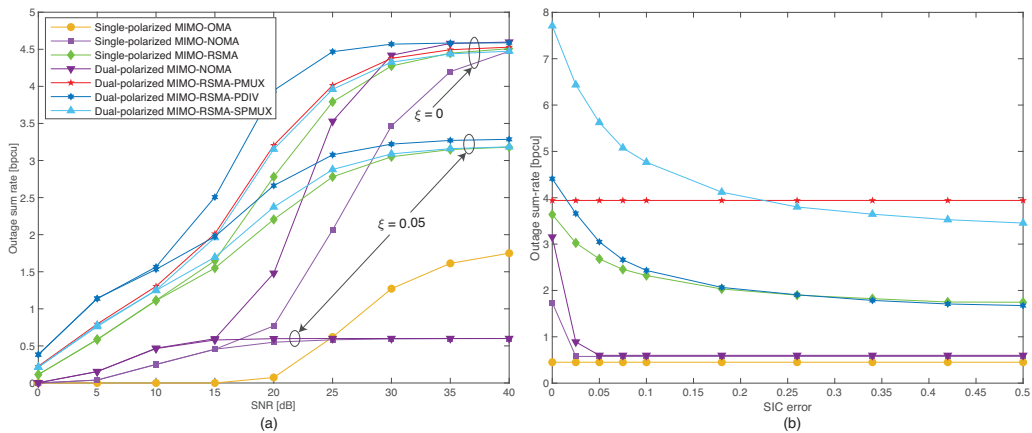}
	\caption{Simulated outage sum-rates for various massive MIMO schemes: (a) outage sum-rate versus SNR with different values of $\xi$, and (b) outage sum-rate versus $\xi$ for a fixed SNR value of $24$~dB ($R^c = 0.5, R^p_{1} = 0.1, R^p_{2} = 1, R^p_{3} = 2$~bpcu, $\chi = 0.001$).}\label{f4}
\end{figure*}
\begin{figure*}[t]
	\centering
	\includegraphics[width=.75\linewidth]{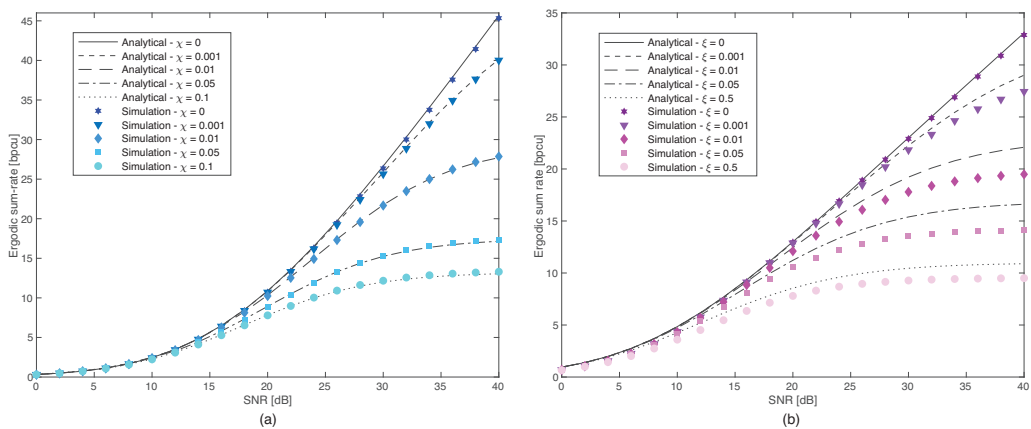}
	\caption{Comparison between analytical and simulated ergodic sum-rates: (a) MIMO-RSMA-PMUX for various values of $\chi$, and (b) MIMO-RSMA-PDIV for various values of $\xi$ with $\chi = 0$.}\label{f5}
\end{figure*}

The analysis for the dual-polarized MIMO-RSMA-SPMUX approach is validated in Figs. \ref{rev_f1}(a) and \ref{rev_f1}(b). We can also observe for this third scheme that the analytical expressions in \eqref{out_comm_a3} and \eqref{out_private_a3} provide accurate approximations of the simulated outage probabilities. Furthermore, these figures suggest that transmitting two simultaneous RSMA streams is not beneficial to improving the reliability of the system, such that the outage probabilities of MIMO-RSMA-SPMUX are noticeably higher than those in Figs. \ref{f2} and \ref{f3}. Such performance behavior is confirmed in Figs. \ref{rev_f2}(a) and \ref{rev_f2}(b), where we put into perspective the outage probabilities of the three proposed dual-polarized MIMO-RSMA schemes. Fig. \ref{rev_f2}(a) considers the case with perfect SIC decoding. As can be seen, in this ideal scenario, the MIMO-RSMA-PDIV is the most reliable scheme for all users in almost the entire SNR range, which is expected since it implements polarization diversity. In the other extreme, the interference introduced in the SIC process makes the MIMO-RSMA-SPMUX the least reliable. In Fig. \ref{rev_f2}(b), SIC decoding errors severely degrade the performance of users $2$ and $3$ when served with MIMO-RSMA-PDIV and MIMO-RSMA-SPMUX. The performance of user $1$ is also degraded. However, since its target rate is low, i.e., $R^p_1 = 0.1$, imperfect SIC does not increase its outage probability substantially. In summary, when the impact of imperfect SIC is severe, the SIC-free MIMO-RSMA-PMUX approach is highly advantageous and provide lower outage probabilities than the two other proposed techniques.

For an RSMA scheme, the outage sum-rate is defined by $\sum_{u = 1}^{U} [(1- P^c_{u}) R^c + (1 - P^p_{u}) R^p_{u}]$, which is the sum of the outage rates for the common and private messages. Figs. \ref{f4}(a) and \ref{f4}(b) present the outage sum-rates achieved by the proposed dual-polarized MIMO-RSMA strategies and by other conventional MA systems. In these examples, for a fair comparison, the target rate for the $u$th user in schemes other than RSMA is adjusted as $R^c + R^p_u$. Specifically, Fig. \ref{f4}(a) reveals the performance superiority of the proposed dual-polarized MIMO-RSMA schemes over the baseline systems in scenarios with and without residual SIC errors. In Fig. \ref{f4}(a), the outage sum-rate curves for the MIMO-RSMA-SPMUX scheme correspond to the data streams received in the vertical polarization. As we can see, in the case with $\xi = 0$, the outage sum-rate performance per polarization of the MIMO-RSMA-SPMUX is close to the MIMO-RSMA-PMUX counterpart. This implies that, under perfect SIC, the sum of the rates achieved in the two polarizations by the MIMO-RSMA-SPMUX scheme should be approximately twice that achieved by the proposed SIC-free approach. This behavior is demonstrated in Fig. \ref{f4}(b), where we plot the outage sum-rates versus SIC error for various MA schemes, but for the MIMO-RSMA-SPMUX in particular, we consider its net outage sum-rate achieved from both vertical and horizontal polarizations. For instance, when $\xi = 0$, the MIMO-RSMA-SPMUX attains almost $8$~bpcu against approximately $4$~bpcu from the MIMO-RSMA-PMUX counterpart. However, when the SIC error factor increases, the outage sum-rates of all SIC-based schemes are degraded, such that, for values of $\xi$ above $0.2$, the MIMO-RSMA-SPMUX is outperformed by the MIMO-RSMA-PMUX approach. This means that, if SIC decoding errors are significant and the outage sum-rate is a priority, the MIMO-RSMA-PMUX should be preferred. On the other hand, among the three proposed MA schemes, it is clear that the MIMO-RSMA-PDIV performs worst in terms of outage sum-rate under imperfect SIC. Nevertheless, we can observe in Figs. \ref{f4}(a) and \ref{f4}(b) that, despite the value of the SIC error, all RSMA schemes (even the single-polarized MIMO-RSMA) can remarkably outperform the baseline counterparts.

The accuracy of the analytical ergodic sum-rate expressions derived in propositions V to VIII is evaluated in Fig. \ref{f5}. More specifically, Fig. \ref{f5}(a) presents the ergodic sum-rates achieved with the MIMO-RSMA-PMUX for different values of $\chi$. As can be observed, a near-perfect agreement between analytical (generated with \eqref{erg_comm_a1} and \eqref{erg_pri_a1}) and simulated curves can be obtained for all values of $\chi$. This result also confirms that cross-polar interference is detrimental to the performance of the proposed dual-polarized approach, with its ergodic sum-rate being limited when $\chi>0$. In Fig. \ref{f5}(b), we compare analytical and simulated ergodic sum-rates achieved with MIMO-RSMA-PDIV for various values of residual SIC errors $\xi$. As can be seen, the approximate curves generated with \eqref{erg_pri_a2} and \eqref{erg_comm_a2} provide upper bounds to the simulated ergodic sum-rates, which are useful for determining the fundamental performance limits of the proposed approach. The harmful effect of imperfect SIC on the ergodic sum-rate performance of the second dual-polarized approach also becomes evident in this result.

\begin{figure*}[t]
	\centering
	\includegraphics[width=.75\linewidth]{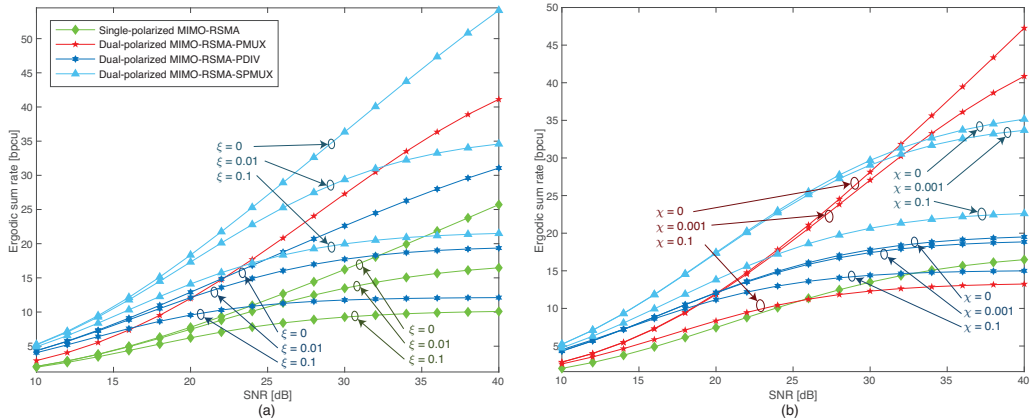}
	\caption{Simulated ergodic sum-rates for different MIMO-RSMA schemes: (a) for various values of $\xi$ with $\chi = 0.001$, and (b) for various values of $\chi$ with $\xi = 0.01$.}\label{f6}
\end{figure*}

\begin{figure*}[t]
	\centering
	\includegraphics[width=1\linewidth]{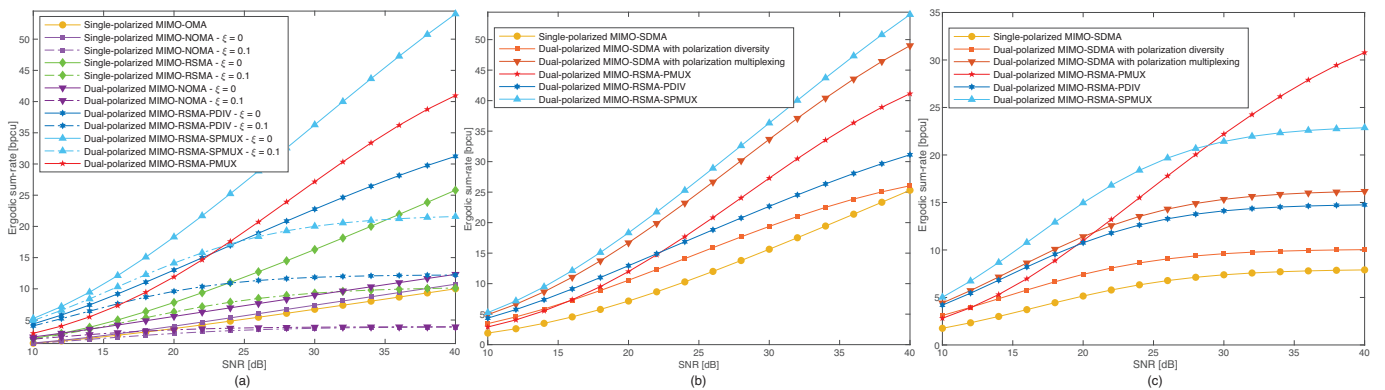}
	\caption{Simulated ergodic sum-rates for various MA schemes: (a) with different values of $\xi$, (b) for $\xi=0$, and (c) for $\xi=0.01$ considering imperfect CSI with a channel error variance of $0.3$ ($\chi = 0.001$).}\label{f7}
\end{figure*}

In Fig. \ref{f6}(a), the ergodic sum-rates of single and dual-polarized MIMO-RSMA schemes are put into perspective for different values of residual SIC errors and considering an iXPD of $\chi = 0.001$. It can be noticed that the three dual-polarized schemes remarkably outperform the single-polarized counterpart for all values $\xi$. We can also observe that the strategy of employing conventional RSMA to each polarization and transmitting two parallel pairs of common and private streams provides ergodic sum-rate gains to the proposed MIMO-RSMA-SPMUX scheme, which can remarkably outperform all the other systems in the scenario with perfect SIC. On the other hand, when SIC errors are present, the SIC-free MIMO-RSMA-PMUX approach outperforms the single-polarized MIMO-RSMA and the dual-polarized MIMO-RSMA-PDIV and MIMO-RSMA-SPMUX schemes in the high-SNR regime. Nevertheless, for low SNR values, MIMO-RSMA-PMUX is still outperformed by the two other dual-polarized systems, even when the residual SIC error is high. 
In Fig. \ref{f6}(b), we set the SIC error factor to $\xi = 0.01$ and investigate the impact of different levels of iXPD on the ergodic sum-rates of the MIMO-RSMA systems. Because the single-polarized MIMO-RSMA does not experience cross-polar interference, its ergodic sum-rate keeps unaltered despite the value of $\chi$. As a result, when the iXPD is high, e.g., when $\chi = 0.1$, in SNR values above $26$~dB for the MIMO-RSMA-PMUX and above $32$~dB for the MIMO-RSMA-PDIV, the single-polarized MIMO-RSMA achieves the best performance. In contrast, despite the data rate degradation, the dual-polarized MIMO-RSMA-SPMUX can still outperform all considered schemes even in the extreme case with $\chi = 0.1$.

Fig. \ref{f7} presents simulated ergodic sum-rates under an iXPD of $\chi = 0.001$. Fig. \ref{f7}(a) confirms again that MIMO-RSMA-SPMUX is the best option in terms of ergodic sum-rate for the ideal case with perfect SIC decoding and scenarios with imperfect SIC in low SNR values. On the other hand, the MIMO-RSMA-PMUX is advantageous mainly in the high-SNR regime under imperfect SIC, whereas MIMO-RSMA-PDIV outperforms the MIMO-RSMA-PMUX scheme in low SNR values but always experiences lower sum-rates than the MIMO-RSMA-SPMUX. It is also confirmed that the proposed dual-polarized MIMO-RSMA approaches can outperform all baseline MA schemes considered in this figure. For instance, for an SNR value of $26$~dB, in the scenario with perfect SIC, the dual-polarized schemes MIMO-RSMA-SPMUX, MIMO-RSMA-PMUX, and MIMO-RSMA-PDIV can achieve remarkable sum-rates of $28.8$~bpcu, $19.14$~bpcu, and $17.41$~bpcu, respectively, against only $7.29$~bpcu observed for the dual-polarized MIMO-NOMA. For $\xi = 0.1$, the sum-rates of the single-polarized MIMO-RSMA, MIMO-RSMA-PDIV, and MIMO-RSMA-SPMUX are degraded. However, even under high interference, the MIMO-RSMA schemes can outperform the conventional systems.

Last, Figs. \ref{f7}(b) and \ref{f7}(c) compare the ergodic sum-rates of the proposed schemes with single and dual-polarized MIMO-SDMA systems. Specifically, assuming perfect SIC, Fig. \ref{f7}(b) shows that MIMO-RSMA-SPMUX achieves the best sum-rate performance even over the dual-polarized MIMO-SDMA scheme with polarization multiplexing. Moreover, despite not attaining the highest sum-rates, the MIMO-RSMA-PDIV and MIMO-RSMA-PMUX schemes can still outperform the dual-polarized MIMO-SDMA with polarization diversity and single-polarized MIMO-SDMA system. To further demonstrate the advantages and robustness of our dual-polarized transmission strategies, Fig. \ref{f7}(c) plots the ergodic sum-rates under both imperfect SIC and imperfect CSI. On the one hand, we can notice that all MIMO-SDMA schemes experience accentuated performance degradation. On the other hand, the dual-polarized MIMO-RSMA schemes show more robustness also to imperfect CSI. For instance, in low SNR values, where the interference from imperfect CSI is not dominant, the MIMO-SDMA schemes can still achieve sum-rates close to the observed in the perfect CSI case, which are superior to those of MIMO-RSMA-PMUX and MIMO-RSMA-PDIV. Nevertheless, from $20$~dB onward, despite conveying single common and private streams, the MIMO-RSMA-PMUX without SIC can remarkably outperform even the MIMO-SDMA with polarization multiplexing. Note that these performance gains have been achieved under a fixed power allocation policy. It is worth highlighting that higher DoFs can be unleashed by the RSMA-based schemes with more advanced adaptive power allocation protocols \cite{Clerckx16, Piovano17}.The results in Fig. \ref{f7} also made clear that depending on the SNR regime, the observed values of iXPD, SIC error factor, and imperfect CSI, different schemes deliver the highest sum-rates. Such behavior suggests that its possible to further improve the performance of dual-polarized MIMO-RSMA systems by smartly shifting between the proposed transmission schemes.

\section{Conclusions}
We have proposed and investigated three RSMA-based transmission strategies for dual-polarized massive MIMO networks. First, MIMO-RSMA-PMUX freed the system from residual SIC errors by multiplexing common and private messages in the polarization domain. Second, MIMO-RSMA-PDIV enabled users to exploit polarization diversity by transmitting replicas of the data messages in the two polarizations. And third, MIMO-RSMA-SPMUX allowed users to experience significant rate gains by applying conventional RSMA to each polarization separately. Simulation results validated the theoretical analysis and showed that the dual-polarized MIMO-RSMA schemes offer attractive features and a trade-off between reliability and throughput. For instance, it was shown that MIMO-RSMA-PDIV achieves the lowest outage probabilities as long as the level of SIC decoding error is sufficiently low. We revealed that MIMO-RSMA-PMUX is advantageous mainly in scenarios with accentuated interference and high SNR values in terms of outage probability, outage sum-rates, and ergodic sum-rates. In its turn, the MIMO-RSMA-SPMUX scheme achieved the worst outage probabilities but could deliver the highest ergodic and outage sum-rates for low SIC error factors and small SNR values. 

% \clearpage
% \newpage

\appendices

\section{Proof of Proposition I}\label{ap1}
\renewcommand{\theequation}{A-\arabic{equation}}
\setcounter{equation}{0}

By recalling the SINR expression from \eqref{sinr_a1_c}, we can rewrite the outage probability for the common message of MIMO-RSMA-PMUX as follows
\begin{align}\label{prob_interm}
    P^c_{gu} &= \mathrm{Pr}\left\{ \varsigma^c_{gu} < \tau_g^c( \omega^c_{gu} + \sigma^2) \right\} \nonumber\\
    &= \int_{0}^{\infty} \int_{0}^{\tau_g^c(y + \sigma^2)} f_{\varsigma^c_{gu} \omega^c_{gu}}(x,y) dx dy,
\end{align}
where we have defined $\tau_g^c = 2^{R^{c}_{g}} - 1$, and $f_{\varsigma^c_{gu} \omega^c_{gu}}(x,y)$ denotes the joint PDF of $\varsigma^c_{gu}$ and $\omega^c_{gu}$. Given that $\varsigma^c_{gu}$ and $\omega^c_{gu}$ are correlated, deriving their exact joint PDF leads to an intractable mathematical analysis. As an alternative, we assume that $\varsigma^c_{gu}$ and $\omega^c_{gu}$ are independent RVs and denote their marginal PDFs by $f_{\varsigma^c_{gu}}(x)$ and $f_{\omega^c_{gu}}(y)$, respectively. From Section \ref{statc_a1}, we know that $\varsigma^c_{gu}$ follows a Gamma distribution with shape parameter $1$ and rate parameter $\phi/\zeta_{gu}\alpha_{g}$, and $\omega^c_{gu}$ follows a Gamma distribution with shape parameter $U$ and rate parameter $\phi/\zeta_{gu}\chi \beta_{gu}$. Therefore, by recalling  \cite[eq. (15.1.1)]{Krishnamoorthy2006}, the following is achieved
\begin{align}\label{deriv1}
    P^c_{gu} &= \frac{\phi^{U} }{\chi^U\Gamma(U) \zeta_{gu}^{U}\beta_{gu}^U}  \left( \int_{0}^{\infty} y^{U-1} e^{-y\frac{\phi}{ \zeta_{gu} \chi \beta_{gu}}} dy  - e^{-\frac{\phi \tau_g^c \sigma^2}{\zeta_{gu} \alpha_{g}}} \right. \nonumber\\
    &\left. \times \int_{0}^{\infty} y^{U-1} e^{-y\left(\frac{\phi}{ \zeta_{gu} \chi \beta_{gu}} + \frac{\phi \tau_g^c}{\zeta_{gu} \alpha_{g}} \right)}  dy \right).
\end{align}
The two integrals in \eqref{deriv1} are of the form $\int_{0}^{\infty} y^{a-1} e^{-yb} dy$, which has a solution given by $b^{-a}\Gamma(a)$ \cite[eq. (3.381.4)]{ref8}. Therefore, by defining the SNR by $\rho = \sigma_n^{-2}$, \eqref{deriv1} can be finally solved as
\begin{align}\label{pout_ap1}
    P^c_{gu} &= \text{\footnotesize $1 - \left(\frac{\alpha_{g}}{\alpha_{g} + \chi\beta_{gu}\tau_g^c } \right)^U  e^{-\frac{\phi \tau_g^c}{\rho \zeta_{gu} \alpha_{g}}}$},
\end{align}
which completes the proof. \hfill $\blacksquare$

\begin{figure*}[!t]
	% ensure that we have normalsize text
	\normalsize
	% Store the current equation number.
	%\setcounter{MYtempeqncnt}{\value{equation}}
	% Set the equation number to one less than the one
	% desired for the first equation here.
	% The value here will have to changed if equations
	% are added or removed prior to the place these
	% equations are referenced in the main text.
% 	\setcounter{equation}{28}

\begin{align}
    P^c_{gu} & = \frac{2\phi}{(1-\chi)^{U+1} \Gamma(U) \zeta_{gu} \beta_{gu}}  \int_{0}^{\infty} e^{-y \frac{\phi}{\zeta_{gu} \beta_{gu}}} \gamma\left(U, y\phi \left[ (\zeta_{gu} \chi \beta_{gu})^{-1} - (\zeta_{gu} \beta_{gu})^{-1} \right]\right) \left[  \frac{1 - \chi}{2} + \chi e^{-\frac{\phi \tau_g^c \sigma^2}{ \zeta_{gu} \chi \alpha_{g}}} e^{-y \frac{\phi \tau_g^c}{ \zeta_{gu} \chi \alpha_{g}}}  \right. \nonumber\\
    & \left.  - e^{-\frac{\phi \tau_g^c \sigma^2}{\zeta_{gu} \alpha_{g}}} e^{-y \frac{\phi \tau_g^c}{\zeta_{gu} \alpha_{g}}} + \frac{\chi^2 e^{-\frac{2 \phi \tau_g^c \sigma^2}{\zeta_{gu} \chi \alpha_{g}}}}{2(1-\chi)}  e^{-y \frac{2 \phi \tau_g^c}{\zeta_{gu} \chi \alpha_{g} }} + \frac{e^{-\frac{2 \phi \tau_g^c \sigma^2}{\zeta_{gu} \alpha_{g}}}}{2(1-\chi)}  e^{-y \frac{2 \phi \tau_g^c}{\zeta_{gu} \alpha_{g}}} - \frac{\chi e^{-\frac{\phi \tau_g^c \sigma^2 (1 + \chi)}{\zeta_{gu} \chi \alpha_{g}}}}{1-\chi}  e^{-y \frac{\phi \tau_g^c (1 + \chi)}{\zeta_{gu} \chi \alpha_{g}} } \right] dy. \tag{C-1}\label{prop_III_01}\\
    P^c_{gu} & = \frac{2\chi}{(1-\chi)^{U+1} \Gamma(U)} \left[ \frac{1 - \chi}{2} \int_{0}^{\infty} e^{-z \frac{\chi}{1-\chi}} \gamma(U, z) dz
    + \chi e^{-\frac{\phi \tau_g^c \sigma^2}{ \zeta_{gu} \chi \alpha_{g}}} 
     \int_{0}^{\infty} e^{-z \frac{\beta_{gu} \tau_g^c + \chi \alpha_{g} }{ (1 - \chi) \alpha_{g}}} \gamma(U, z) dz
    \right. \nonumber\\
     &  - e^{-\frac{\phi \tau_g^c \sigma^2}{\zeta_{gu} \alpha_{g}}} 
     \int_{0}^{\infty} e^{-z \frac{\chi \beta_{gu}\tau_g^c + \chi \alpha_{g} }{ (1 - \chi) \alpha_{g}}} \gamma(U, z) dz
      + \frac{\chi^2 e^{-\frac{2 \phi \tau_g^c \sigma^2}{\zeta_{gu} \chi \alpha_{g}}}}{2(1-\chi)}
     \int_{0}^{\infty} e^{-z \frac{ 2 \beta_{gu} \tau_g^c + \chi \alpha_{g} }{ (1 - \chi) \alpha_{g}}} \gamma(U, z) dz \nonumber\\
    & \left. + \frac{e^{-\frac{2 \phi \tau_g^c \sigma^2}{\zeta_{gu} \alpha_{g}}}}{2(1-\chi)} 
    \int_{0}^{\infty} e^{-z \frac{ 2 \chi \beta_{gu} \tau_g^c + \chi \alpha_{g} }{ (1 - \chi) \alpha_{g}}} \gamma(U, z) dz 
        -\frac{\chi e^{-\frac{\phi \tau_g^c \sigma^2 (1 + \chi)}{\zeta_{gu} \chi \alpha_{g}}}}{1-\chi}
    \int_{0}^{\infty} e^{-z \frac{ (1 + \chi) \beta_{gu} \tau_g^c + \chi \alpha_{g} }{ (1 - \chi) \alpha_{g}}} \gamma(U, z) dz \right] dy. \tag{C-2}\label{ints_ap4}
\end{align}

	% Restore the current equation number.
% 	\setcounter{equation}{25}
	% The IEEE uses as a separator
	\hrulefill
	% The spacer can be tweaked to stop underfull vboxes.
\end{figure*}

\section{Proof of Proposition II}\label{ap2}
\renewcommand{\theequation}{B-\arabic{equation}}
\setcounter{equation}{0}

Recall from Section \ref{statc_a1} that $\varsigma^p_{gu}$ and $\omega^p_{gu}$ follow Gamma distributions with shape parameters 1 and rate parameters $\phi / \zeta_{gu} \beta_{gu}$ and $\phi/\zeta_{gu}\chi\alpha_{g}$, respectively. Given this, and assuming that $\varsigma^p_{gu}$ and $\omega^p_{gu}$ are independent, their joint PDF can be obtained by
\begin{align}\label{jointpdf2}
    f_{\varsigma^p_{gu} \omega^p_{gu}}(x,y) &= f_{\varsigma^p_{gu}}(x)f_{\omega^p_{gu}}(y) \nonumber\\
    &= \frac{\phi^2}{\zeta_{gu}^2 \chi\alpha_{g} \beta_{gu}} e^{-y\frac{\phi}{\zeta_{gu}\chi\alpha_{g}}} e^{-x\frac{\phi}{\zeta_{gu} \beta_{gu}}},
\end{align}
where $f_{\varsigma^p_{gu}}(x)$ and $f_{\omega^p_{gu}}(y)$ represents the marginal distributions of $\varsigma^p_{gu}$ and $\omega^{p}_{gu}$.

Then, with the SINR in \eqref{sinr_a1_p} and defining $\tau_{gu}^p = 2^{R^{p}_{gu}} - 1$, the outage probability for the private massages with MIMO-RSMA-PMUX can be derived as follows
\begin{align}\label{probp2}
    P^p_{gu} & = \frac{\phi}{\zeta_{gu}\chi\alpha_{g}}  \left( \int_{0}^{\infty}  e^{-y\frac{\phi}{\zeta_{gu}\chi\alpha_{g}}} dy - e^{-\frac{\phi}{\zeta_{gu} \beta_{gu}} \tau_{gu}^p\sigma^2} \right.\nonumber\\
    &\times \left. \int_{0}^{\infty} e^{-y\left( \frac{\phi}{\zeta_{gu} \beta_{gu}} \tau_{gu}^p + \frac{\phi}{\zeta_{gu}\chi\alpha_{g}} \right)}  dy \right) \nonumber\\
    &= 1 - \frac{\beta_{gu}}{ \chi\alpha_{g}\tau_{gu}^p + \beta_{gu}} e^{-\frac{\phi \tau_{gu}^p}{\rho\zeta_{gu}\beta_{gu}}},
\end{align}
which completes the proof. \hfill $\blacksquare$

\section{Proof of Proposition III}\label{ap4}
\renewcommand{\theequation}{C-\arabic{equation}}
\setcounter{equation}{1}
As discussed in Section \ref{statc_a2}, the RVs $\varrho^{c,\star}_{gu}$ and $\varpi^{c,\star}_{gu}$ are correlated, which complicates the derivation of their joint PDF. Therefore, for MIMO-RSMA-PDIV, we also assume that $\varrho^{c,\star}_{gu}$ and $\varpi^{c,\star}_{gu}$ are independent. Under this assumption, the desired joint PDF can be obtained through the product between the marginal PDFs of $\varrho^{c,\star}_{gu}$ and $\varpi^{c,\star}_{gu}$, and we can achieve \eqref{prop_III_01}.
Then, we apply the change of variables $y\phi \left[ (\zeta_{gu} \chi \beta_{gu})^{-1} - (\zeta_{gu} \beta_{gu})^{-1} \right] = z$, which results in \eqref{ints_ap4}, where \eqref{prop_III_01} and \eqref{ints_ap4} are shown on the top of this page.

Next, we solve the integrals in \eqref{ints_ap4} by using \cite[eq. (6.451.1)]{ref8}. Then, after performing some algebraic manipulations, the outage probability expression for the common message of MIMO-RSMA-PDIV can be finally obtained as in \eqref{out_comm_a2}, which completes the proof. \hfill $\blacksquare$

\section{Proof of Proposition IV}\label{ap5}
\renewcommand{\theequation}{D-\arabic{equation}}
\setcounter{equation}{0}
The gains $\varrho^{p,\star}_{gu}$ and $\varpi^{p,\star}_{gu}$ are also assumed to be independent. Then, by recalling the marginal PDFs of $\varrho^{p,\star}_{gu}$ and $\varpi^{p,\star}_{gu}$ derived in section \ref{statc_a2}, the expression in \eqref{apppa23}, shown on the top of the next page, can be achieved.

By applying results from \cite[eq. (6.451.1)]{ref8} to solve each term of the integrals in \eqref{apppa23} and performing some algebraic manipulations, the outage probability for the private messages of MIMO-RSMA-PDIV can be finally solved as in \eqref{out_private_a2}, which completes the proof. \hfill $\blacksquare$

\section{Proof of Proposition VII}\label{ap6}
\renewcommand{\theequation}{E-\arabic{equation}}
\setcounter{equation}{0}
First, we need to obtain the CDF of $\min \{\gamma^c_{gu}\}$, for $u=1, \cdots, U$, denoted by $F_{\hspace{-1mm}\text{\tiny min$\gamma^c_{gu}$}}(z)$. From Appendix \ref{ap1}, we can easily achieve the CDF of $\gamma^c_{gu}$ by replacing $\tau^c_{gu}$ by $z$ in the outage probability expression in \eqref{pout_ap1}. In view of this, we can exploit the theory of Order Statistics to calculate $F_{\hspace{-1mm}\text{\tiny min$\gamma^c_{gu}$}}(z)$. Specifically, by knowing that $\gamma^c_{g1}, \cdots, \gamma^c_{gU}$ are not identically distributed, we can recall \cite[eq. (5.4.11)]{ref7} to obtain the desired CDF as follows
\begin{align}
    F_{\hspace{-1mm}\text{\tiny min$\gamma^c_{gu}$}}(z) &= 1 - \prod_{l = 1}^{U} \left(\frac{\alpha_{g}}{\alpha_{g} + \chi\beta_{gu}z } \right)^U  e^{-z\frac{\phi}{\rho\zeta_{gl}\alpha_{g}}} \nonumber\\
    &= 1 - \alpha_{g}^{U^2} \left({\alpha_{g} + \chi\beta_{gu}z } \right)^{-U^2}  e^{-z\frac{\phi \sum_{l=1}^{U} \frac{1}{\zeta_{gl}}}{\rho\alpha_{g}}},
\end{align}
Now that we know $F_{\hspace{-1mm}\text{\tiny min$\gamma^c_{gu}$}}(z)$, the desired sum-rate can be obtained by a Riemann-Stieltjes integral, as follows
\begin{align}
    C^c_{g} & = \sum_{u=1}^{U} \frac{\alpha^{U^2}_{g}}{\mathrm{ln}(2)} \int_{0}^{\infty}  (1 + z)^{-1}(\alpha_{g} + \chi\beta_{gu}z)^{-U^2} \nonumber\\
    &\times e^{-z\frac{\phi \sum_{l=1}^{U} \frac{1}{\zeta_{gl}}}{\rho\alpha_{g}}} dz.
\end{align}
Next, by integrating by parts and applying the transformation $\alpha_{g} + \chi\beta_{gu} z = t$, we achieve
\begin{align}\label{preerg_ap6}
    C^c_{g} & = \sum_{u=1}^{U} \frac{\alpha^{U^2}_{g}}{\mathrm{ln}(2)} \left[ - \alpha_{g}^{-U^2} e^{\frac{\phi \sum_{l=1}^{U} \frac{1}{\zeta_{gl}} }{\rho \alpha_{g}} }  \right. \nonumber\\
    &\times \mathrm{Ei}\left(- \frac{\phi \sum_{l=1}^{U} \frac{1}{\zeta_{gl}} }{\rho \alpha_{g}} \right) + U^2 e^{\frac{\phi \sum_{l=1}^{U} \frac{1}{\zeta_{gl}} }{\rho \alpha_{g}} } \int_{\alpha_{g}}^{\infty}  t^{-U^2 - 1}\nonumber\\
    &\times \left. \mathrm{Ei}\left(-\left(\frac{\chi\beta_{gu} - \alpha_{g}}{\chi\beta_{gu}} + \frac{t}{\chi\beta_{gu}} \right)\frac{\phi \sum_{l=1}^{U} \frac{1}{\zeta_{gl}}}{\rho\alpha_{g}} \right) dt \right].
\end{align}

For solving the integral in \eqref{preerg_ap6}, we make use of \cite[eq. (4.1.24)]{Geller69}, in which, after some simplifications, we finally obtain the desired ergodic sum-rate expression, as in \eqref{prop_vii_01}, shown on the top of the next page, which completes the proof. \hfill $\blacksquare$

\begin{figure*}[!t]
	% ensure that we have normalsize text
	\normalsize
	% Store the current equation number.
	%\setcounter{MYtempeqncnt}{\value{equation}}
	% Set the equation number to one less than the one
	% desired for the first equation here.
	% The value here will have to changed if equations
	% are added or removed prior to the place these
	% equations are referenced in the main text.
% 	\setcounter{equation}{28}

\begin{align}
    P^p_{gu} &= \frac{2(\xi \alpha_{g})^{U-1}}{\Gamma(U-1)} \left( (\xi \alpha_{g} - \beta_{gu})^{-U} \int_{0}^{\infty} \left[ -\frac{\chi \beta_{gu}}{2 (1 - \chi) } e^{-z \frac{\beta_{gu}}{\xi \alpha_{g} - \beta_{gu}}} \gamma(U-1, z) 
    -  \frac{\chi^2 \beta_{gu} e^{-\frac{\phi \tau_{gu}^p\sigma^2}{\zeta_{gu} \chi \beta_{gu}}} }{ (1 - \chi)^2 } e^{-z \frac{ \beta + \xi \alpha_{g} \tau_{gu}^p }{\xi \alpha_{g} - \beta_{gu}}} \gamma(U-1, z)
    \right. \right. \nonumber\\
    & + \frac{\chi \beta_{gu} e^{-\frac{\phi \tau_{gu}^p\sigma^2}{\zeta_{gu} \beta_{gu}}} }{(1 - \chi)^2 } e^{-z \frac{ \beta_{gu} + \xi \chi \alpha_{g} \tau_{gu}^p }{\xi \alpha_{g} - \beta_{gu}}} \gamma(U-1, z) - \frac{\chi^3 \beta_{gu} e^{-\frac{2\phi \tau_{gu}^p\sigma^2}{\zeta_{gu} \chi \beta_{gu}}} }{2 (1 - \chi)^3} e^{-z \frac{\beta_{gu} + 2 \xi \alpha_{g} \tau_{gu}^p }{\xi \alpha_{g} - \beta_{gu}}} \gamma(U-1, z) \nonumber\\
    &\left.  - \frac{\chi \beta_{gu} e^{-\frac{2 \phi \tau_{gu}^p\sigma^2}{\zeta_{gu} \beta_{gu}}} }{2 (1 - \chi)^3} e^{-z \frac{ \beta_{gu} + 2 \xi\chi \alpha_{g} \tau_{gu}^p }{\xi \alpha_{g} - \beta_{gu}}} \gamma(U-1, z) +  \frac{\chi^2 \beta_{gu} e^{-\frac{\phi \tau_{gu}^p\sigma^2 (1 + \chi)}{\zeta_{gu} \chi \beta_{gu}}} }{ (1 - \chi)^3} e^{-z \frac{\chi \beta_{gu} + \xi\alpha_{g} \tau_{gu}^p \left(1 + \chi \right) }{\chi}} \gamma(U-1, z) \right] dz \nonumber\\
    &- (\xi \alpha_{g} - \chi \beta_{gu})^{-U}  \int_{0}^{\infty} \left[ -\frac{\chi \beta_{gu}}{2 (1 - \chi) } e^{-w \frac{\beta_{gu}}{\xi \alpha_{g} - \chi \beta_{gu}}} \gamma(U-1, w)  -  \frac{\chi^2 \beta_{gu} e^{-\frac{\phi \tau_{gu}^p\sigma^2}{\zeta_{gu} \chi \beta_{gu}}} }{ (1 - \chi)^2 } e^{-w \frac{ \beta + \xi \alpha_{g} \tau_{gu}^p }{\xi \alpha_{g} - \chi \beta_{gu}}} \gamma(U-1, w) \right. \nonumber\\
    &+ \frac{\chi \beta_{gu} e^{-\frac{\phi \tau_{gu}^p\sigma^2}{\zeta_{gu} \beta_{gu}}} }{(1 - \chi)^2 } e^{-w \frac{ \beta_{gu} + \xi \chi \alpha_{g} \tau_{gu}^p }{\xi \alpha_{g} - \chi \beta_{gu}}} \gamma(U-1, w) - \frac{\chi^3 \beta_{gu} e^{-\frac{2\phi \tau_{gu}^p\sigma^2}{\zeta_{gu} \chi \beta_{gu}}} }{2 (1 - \chi)^3} e^{-w \frac{\beta_{gu} + 2 \xi \alpha_{g} \tau_{gu}^p }{\xi \alpha_{g} - \chi \beta_{gu}}} \gamma(U-1, w) \nonumber\\
    & \left. \left. - \frac{\chi \beta_{gu} e^{-\frac{2 \phi \tau_{gu}^p\sigma^2}{\zeta_{gu} \beta_{gu}}} }{2 (1 - \chi)^3} e^{-w \frac{ \beta_{gu} + 2 \xi\chi \alpha_{g} \tau_{gu}^p }{\xi \alpha_{g} - \chi \beta_{gu}}} \gamma(U-1, w)  +  \frac{\chi^2 \beta_{gu} e^{-\frac{\phi \tau_{gu}^p\sigma^2 (1 + \chi)}{\zeta_{gu} \chi \beta_{gu}}} }{ (1 - \chi)^3} e^{-w \frac{\chi \beta_{gu} + \xi\alpha_{g} \tau_{gu}^p \left(1 + \chi \right) }{\chi}} \gamma(U-1, w) \right] dw \right). \tag{D-1}\label{apppa23}
\end{align}

	% Restore the current equation number.
% 	\setcounter{equation}{25}
	% The IEEE uses as a separator
	\hrulefill
	% The spacer can be tweaked to stop underfull vboxes.
\end{figure*}

\section{Proof of Proposition VIII}\label{ap7}
\renewcommand{\theequation}{F-\arabic{equation}}
\setcounter{equation}{0}
For deriving the ergodic sum-rate of the private messages for MIMO-RSMA-PMUX, we obtain the CDF of $\gamma^p_{gu}$ from the outage probability in \eqref{probp2}. Then, similarly as in Appendix \ref{ap6}, the desired sum-rate is calculated by a Riemann-Stieltjes integral, as follows 
\begin{align}
    C^p_{g} & = \sum_{u=1}^{U} \frac{\beta_{gu}}{\mathrm{ln}(2)}\int_{0}^{\infty} (1 + z)^{-1}(\chi\alpha_{g}z + \beta_{gu})^{-1}\nonumber\\
    &\times e^{-z\frac{\phi}{\rho\zeta_{gu}\beta_{gu}}} dz.
\end{align}
By integrating by parts and applying the transformation $\chi\alpha_{g}z + \beta_{gu} = t$, we get
\begin{align}
    C^p_{g} &= \sum_{u=1}^{U}  \frac{\beta_{gu}}{\mathrm{ln}(2)}  \left[ \beta_{gu}^{-1} e^{\frac{\phi}{\rho\zeta_{gu}\beta_{gu}}} \mathrm{Ei}\left( -\frac{\phi}{\rho\zeta_{gu}\beta_{gu}} \right) + e^{\frac{\phi}{\rho\zeta_{gu}\beta_{gu}}} \right. \nonumber\\
     & \times \left. \int_{\beta_{gu}}^{\infty} \hspace{-2mm} t^{-2} \mathrm{Ei}\left( -\left(\frac{t}{\chi \alpha_{g}} + \frac{\chi \alpha_{g} - \beta_{gu}}{\chi \alpha_{g} }  \right) \frac{\phi}{\rho\zeta_{gu}\beta_{gu}} \right) dt \right].
\end{align}

Finally, by recalling \cite[eq. (4.1.21)]{Geller69}, and performing some algebraic manipulations, the ergodic sum-rate of the private messages for the $g$ group can be obtained as
\begin{align}
    C^p_g &=  \sum_{u=1}^{U} \frac{ \beta_{gu}}{ \mathrm{ln}(2)(\beta_{gu} - \chi \alpha_{g})} \left[ e^{\frac{\phi}{\rho \zeta_{gu}\chi\alpha_{g}}} \mathrm{Ei}\left(- \frac{ \phi}{\rho \zeta_{gu}\chi\alpha_{g}} \right)\right. \nonumber\\
    &\left. - e^{\frac{\phi}{\rho \zeta_{gu}\beta_{gu} }} \mathrm{Ei}\left(- \frac{ \phi}{\rho \zeta_{gu} \beta_{gu} } \right)  \right],
\end{align}
which completes the proof. \hfill $\blacksquare$

\begin{figure*}[!t]
	% ensure that we have normalsize text
	\normalsize
	% Store the current equation number.
	%\setcounter{MYtempeqncnt}{\value{equation}}
	% Set the equation number to one less than the one
	% desired for the first equation here.
	% The value here will have to changed if equations
	% are added or removed prior to the place these
	% equations are referenced in the main text.
% 	\setcounter{equation}{28}

\begin{align}
C^c_g &=  \sum_{u=1}^{U} \frac{(-1)^{U^2 - 1}}{\mathrm{ln}(2)}  \left(\frac{\alpha_{g}}{\chi\beta_{gu} - \alpha_{g}} \right)^{U^2} e^{\frac{\phi \sum_{l=1}^{U} \frac{1}{\zeta_{gl}} }{\rho \alpha_{g}} } \left[ \mathrm{Ei}\left(- \frac{\phi \sum_{l=1}^{U} \frac{1}{\zeta_{gl}} }{\rho \alpha_{g}}  \right) \right.%\nonumber\\
    - \bm{e}_{U^2-1}\left( \frac{\phi(\chi \beta_{gu} - \alpha_{g}) \sum_{l=1}^{U} \frac{1}{\zeta_{gl}}}{ \chi \rho \beta_{gu} \alpha_{g}}\right) \nonumber\\
    & \left.\times 
    e^{- \frac{\phi(\chi \beta_{gu} - \alpha_{g}) \sum_{l=1}^{U} \frac{1}{\zeta_{gl}}}{ \chi \rho \beta_{gu} \alpha_{g}} } \mathrm{Ei}\left( - \frac{\phi \sum_{l=1}^{U} \frac{1}{\zeta_{gl}}}{\rho \chi \beta_{gu} }  \right) 
    +  e^{-\frac{\phi \sum_{l=1}^{U} \frac{1}{\zeta_{gl}} }{\rho \alpha_{g}} } \sum_{m = 1}^{U^2 - 1}  \frac{1}{m!} \left(- \frac{\chi \beta_{gu} - \alpha_{g} }{\alpha_{g} } \right)^{m}\right. \nonumber\\
    &\times \left. \sum_{k = 0}^{m-1}  (m-k-1)!  \left( - \frac{\phi \sum_{l=1}^{U} \frac{1}{\zeta_{gl}}}{\rho \chi \beta_{gu} } \right)^K \right].
     \tag{E-4}\label{prop_vii_01}
\end{align}

\begin{align}
    C^p_{g} & = \sum_{u=1}^{U} \left[ \frac{ 2\beta_{gu}^2}{ \mathrm{ln}(2) \chi (1 - \chi)} \int_{0}^{\infty} \frac{ \left(1 + \chi z \right)^{-U+1} e^{-z\frac{\phi}{\rho\zeta_{gu}\beta_{gu} } } }{(1+z)(\xi \alpha_{g} z + \beta_{gu}) (\xi \alpha_{g} z + \chi^{-1} \beta_{gu} )} dz - \frac{ 2\chi^2 \beta_{gu}^2 }{\mathrm{ln}(2) (1 - \chi) }  \right. \nonumber\\
    & \times \int_{0}^{\infty} \frac{\left(1 + z \right)^{-U+1} e^{- z\frac{\phi }{\rho\zeta_{gu}\chi\beta_{gu} } }}{(1+z)(\xi \alpha_{g} z + \beta_{gu})(\xi \alpha_{g} z + \chi\beta_{gu})} dz  - \frac{\chi^3 \beta_{gu}^2 }{\mathrm{ln}(2) (1-\chi)^2 } \int_{0}^{\infty} \frac{\left(1 + 2 z \right)^{-U+1} e^{- z\frac{2\phi }{\rho\zeta_{gu}\chi\beta_{gu} } } }{(1+z)({2\xi \alpha_{g} z + \beta_{gu}}) ({2\xi \alpha_{g} z + \chi \beta_{gu}})} \nonumber\\
    & - \frac{ \beta_{gu}^2 }{  \mathrm{ln}(2) \chi (1-\chi)^2} \int_{0}^{\infty} \frac{\left(1 + 2\chi z \right)^{-U+1} e^{-z \frac{2\phi }{\rho\zeta_{gu}\beta_{gu} } } }{(1+z)({2\xi  \alpha_{g} z + \beta_{gu}}) ({2\xi \alpha_{g} z + \chi^{-1}\beta_{gu} })} dz + \frac{2 \chi^2 \beta_{gu}^2 }{\mathrm{ln}(2) (1-\chi)^2 }  \nonumber\\
    & \left. \times \int_{0}^{\infty} \frac{\left(1 + (1 + \chi) z \right)^{-U+1} e^{- z \frac{\phi(1 + \chi)}{\rho\zeta_{gu}\chi\beta_{gu} } }}{(1+z)(\xi \alpha_{g} z(1 + \chi) + \beta_{gu}) (\xi \alpha_{g} z(1 + \chi) + \chi\beta_{gu})} dz \right].\tag{G-1}\label{erg_p_a2_1}
\end{align}

	% Restore the current equation number.
% 	\setcounter{equation}{25}
	% The IEEE uses as a separator
	\hrulefill
	% The spacer can be tweaked to stop underfull vboxes.
\end{figure*}

\section{Proof of Proposition IX}\label{ap8}
\renewcommand{\theequation}{G-\arabic{equation}}
\setcounter{equation}{1}
Similarly as in the previous appendices, we can obtain the CDF of $\gamma^p_{gu}$ for MIMO-RSMA-PDIV from the outage probability in \eqref{out_private_a2}. Consequently, the ergodic sum-rate for the private messages can be achieved as in \eqref{erg_p_a2_1}, shown on the top of the next page.

Note that all integrals in \eqref{erg_p_a2_1} share a similar form, which can be written as
\begin{align}\label{erg_int_p_a2}
    \Theta(a, b, c, d, h, l) & = \int_{0}^{\infty} \frac{ \left(1 + c z \right)^{-l+1} e^{-z h} }{(1+z)(a z + b) (a z + d b)} dz,
\end{align}
where $a, b, c, d, h$ and $l$ are determined by comparing \eqref{erg_int_p_a2} with the integrals in \eqref{erg_p_a2_1}. By integrating \eqref{erg_int_p_a2} by parts and applying $1 + c z = t$, the following can be achieved
\begin{align}\label{erg_p_a2_2}
    \Theta(\cdot) & =  \frac{1}{b (d - 1)(a - b) (a - d b)} \left[ (a - d b) e^{\frac{b h}{a}} \mathrm{Ei}\left( - \frac{b h}{a} \right) \right. \nonumber\\
    &- (a - b) e^{\frac{b h d }{a}} \mathrm{Ei}\left( - \frac{b h d }{a} \right)
      + (d - 1) b e^{h} \mathrm{Ei}\left( - h \right) \bigg] \nonumber\\
    &  - \frac{l-1}{b (d - 1)(a - b) (a - d b)} \bigg[ (a - b) e^{\frac{b h d}{a}} \nonumber\\
    &\times \int_{1}^{\infty} t^{-l} \mathrm{Ei}\left( - \frac{ h (b c d - a) }{ac} - \frac{ h t }{c} \right) dt - (a - d b) 
       \nonumber \\
    & \times e^{\frac{b h}{a}} \int_{1}^{\infty} t^{-l}  \mathrm{Ei}\left( - \frac{ h (b c - a) }{ac} - \frac{ h t }{c} \right) dt - (d - 1) b e^{h} \nonumber\\
     &\left.  \times \int_{1}^{\infty} t^{-l} \mathrm{Ei}\left( - \frac{ h (c - 1) }{c} - \frac{ h t }{c} \right) dt  \right].
\end{align}

The integrals in \eqref{erg_p_a2_2} are of the form $\int t^{-n} \mathrm{Ei} (\mu t  + \nu)  dt $, which has solution given by \cite[eq. (4.1.24)]{Geller69}, if $\nu\neq 0$, or by \cite[eq. (4.1.23)]{Geller69}, if $\nu= 0$. Then, after solving \eqref{erg_p_a2_2} and replacing into \eqref{erg_p_a2_1}, the ergodic sum-rate expression for the private messages of PDIV can be expressed as in \eqref{erg_pri_a2}, which completes the proof. \hfill $\blacksquare$

\ifCLASSOPTIONcaptionsoff
\newpage
\fi

\bibliographystyle{IEEEtran}
\bibliography{main}

\begin{IEEEbiography}[{\includegraphics[width=1in,height=1.25in,clip,keepaspectratio]{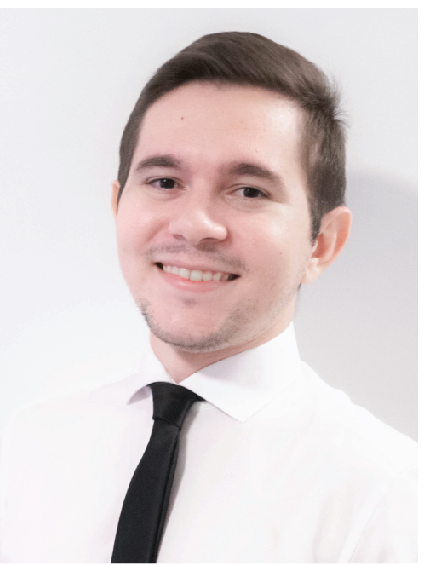}}]{Arthur Sousa de Sena}

is a researcher at the AI and Digital Science Research Center at the Technology Innovation Institute, Abu Dhabi, the United Arab Emirates. Prior to that, from 2019 to 2022, he was a researcher with the Laboratory of Control Engineering and Digital Systems at Lappeenranta-Lahti University of Technology (LUT), Lappeenranta, Finland. He received his D.Sc. degree in Electrical Engineering from LUT, Finland, in 2022, and the M.Sc. and B.Sc. degrees in Teleinformatics Engineering and Computer Engineering from the Federal University of Ceará, Brazil, in 2019 and 2017, respectively. Dr. Sena was awarded the Brazil Scientific Mobility Program scholarship for the period August 2014 to December 2015, in which he had the opportunity to study Computer Engineering as an exchange student at the Armour College of Engineering, Illinois Institute of Technology, Chicago, USA. Due to his outstanding research work in wireless communication, he received the Nokia Foundation Award in October 2020 and the LUT Research Foundation Award in December 2020. Dr. Sena has authored several peer-reviewed papers in prestigious journals and flagship conferences. He is a member of the IEEE and the IEEE Communications Society.

\end{IEEEbiography}

\begin{IEEEbiography}[{\includegraphics[width=1in,height=1.25in,clip,keepaspectratio]{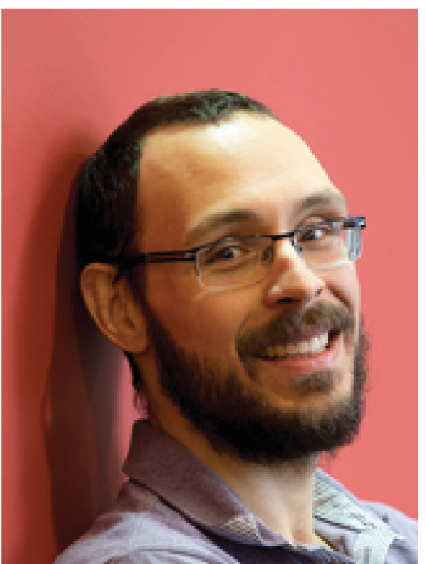}}]{Pedro H. J. Nardelli}

(M'07, SM'19) received the B.S. and M.Sc. degrees in electrical engineering from the State University of Campinas, Brazil, in 2006 and 2008, respectively. In 2013, he received his doctoral degree from University of Oulu, Finland, and State University of Campinas following a dual degree agreement. He is currently Associate Professor (tenure track) in IoT in Energy Systems at LUT University, Finland, and holds a position of Academy of Finland Research Fellow with a project called Building the Energy Internet as a large-scale IoT-based cyber-physical system that manages the energy inventory of distribution grids as discretized packets via machine-type communications (EnergyNet). He leads the Cyber-Physical Systems Group at LUT, and is Project Coordinator of the CHIST-ERA European consortium Framework for the Identification of Rare Events via Machine Learning and IoT Networks (FIREMAN) and of the project Swarming Technology for Reliable and Energy-aware Aerial Missions (STREAM) supported by Jane and Aatos Erkko Foundation. He is also Docent at University of Oulu in the topic of “communications strategies and information processing in energy systems”. His research focuses on wireless communications particularly applied in industrial automation and energy systems. He received a best paper award of IEEE PES Innovative Smart Grid Technologies Latin America 2019 in the track “Big Data and Internet of Things”. He is also IEEE Senior Member. More information: \url{https://sites.google.com/view/nardelli/}

\end{IEEEbiography}

\begin{IEEEbiography}[{\includegraphics[width=1in,height=1.25in,clip,keepaspectratio]{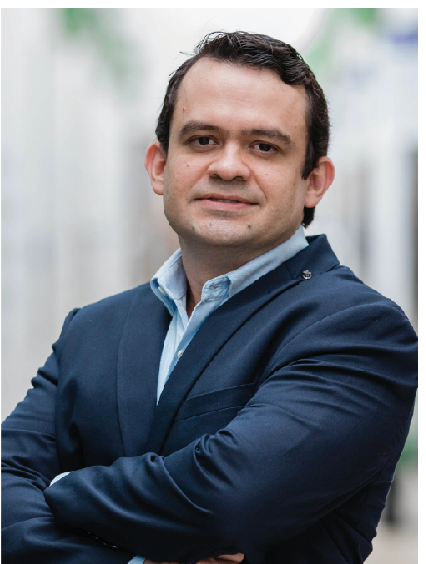}}]{Daniel Benevides da Costa}

(S'04-M'08-SM'14) was born in Fortaleza, Ceará, Brazil, in 1981. He received the B.Sc. degree in Telecommunications from the Military Institute of Engineering (IME), Rio de Janeiro, Brazil, in 2003, and the M.Sc. and Ph.D. degrees in Electrical Engineering, Area: Telecommunications, from the University of Campinas, SP, Brazil, in 2006 and 2008, respectively. His Ph.D. thesis was awarded the Best Ph.D. Thesis in Electrical Engineering by the Brazilian Ministry of Education (CAPES) at the 2009 CAPES Thesis Contest. From 2008 to 2009, he was a Postdoctoral Research Fellow with INRS-EMT, University of Quebec, Montreal, QC, Canada. From 2010 to 2022, he was with the Federal University of Ceará, Brazil. From January 2019 to April 2019, he was Visiting Professor at Lappeenranta University of Technology (LUT), Finland, with financial support from Nokia Foundation. He was awarded with the prestigious Nokia Visiting Professor Grant. From May 2019 to August 2019, he was with King Abdullah University of Science and Technology (KAUST), Saudi Arabia, as a Visiting Faculty, and from September 2019 to November 2019, he was a Visiting Researcher at Istanbul Medipol University, Turkey. From 2021 to 2022, he was Full Professor at the National Yunlin University of Science and Technology (YunTech), Taiwan. Since 2022, he is Principal Researcher of the AI and Digital Science Research Center at the Technology Innovation Institute (TII), a global research center and the applied pillar of Abu Dhabi's Advanced Technology Research Council. He is Editor of several IEEE journals and has acted as Symposium/Track Co-Chair in numerous IEEE flagship conferences.
\end{IEEEbiography}

\begin{IEEEbiography}[{\includegraphics[width=1in,height=1.25in,clip,keepaspectratio]{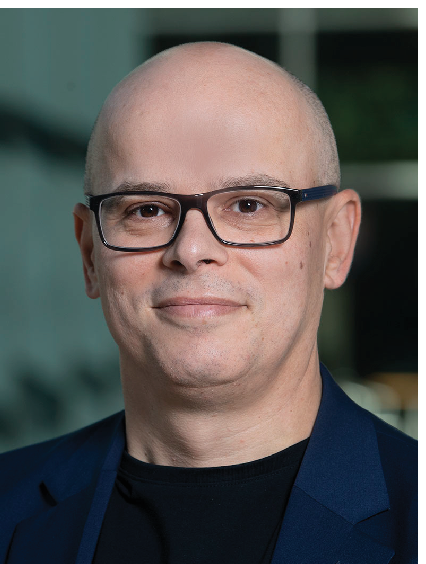}}]{Petar Popovski}

(S'97–A'98–M'04–SM'10–F'16) is a Professor at Aalborg University, where he heads the section on Connectivity and a Visiting Excellence Chair at the University of Bremen. He received his Dipl.-Ing and M. Sc. degrees in communication engineering from the University of Sts. Cyril and Methodius in Skopje and the Ph.D. degree from Aalborg University in 2005. He received an ERC Consolidator Grant (2015), the Danish Elite Researcher award (2016), IEEE Fred W. Ellersick prize (2016), IEEE Stephen O. Rice prize (2018), Technical Achievement Award from the IEEE Technical Committee on Smart Grid Communications (2019), the Danish Telecommunication Prize (2020) and Villum Investigator Grant (2021). He was a Member at Large at the Board of Governors in IEEE Communication Society 2019-2021. He is currently an Editor-in-Chief of IEEEE JOURNAL ON SELECTED AREAS IN COMMUNICATIONS. He also serves as a Vice-Chair of the IEEE Communication Theory Technical Committee and the Steering Committee of IEEE TRANSACTIONS ON GREEN COMMUNICATIONS AND NETWORKING. Prof. Popovski was the General Chair for IEEE SmartGridComm 2018 and IEEE Communication Theory Workshop 2019. His research interests are in the area of wireless communication and communication theory. He authored the book ``Wireless Connectivity: An Intuitive and Fundamental Guide'', published by Wiley in 2020.

\end{IEEEbiography}

\begin{IEEEbiography}[{\includegraphics[width=1in,height=1.25in,clip,keepaspectratio]{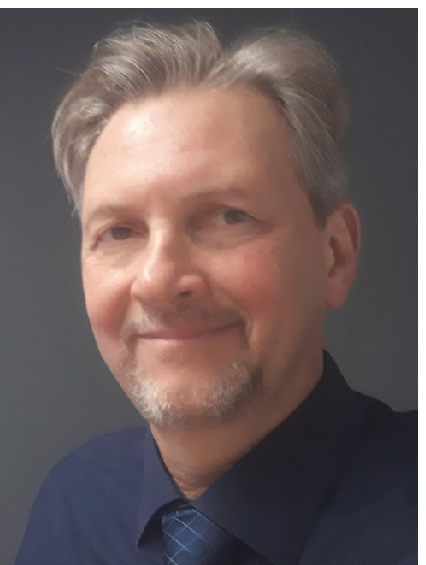}}]{Constantinos B. Papadias}

is the founding Executive Director of the Research, Technology and Innovation Network (RTIN) of The American College of Greece (ACG), in Athens, Greece and the Head of RTIN’s Smart Wireless Future Technologies (SWIFT) Lab, since 2020. He is also Professor of Information Technology at ACG’s Deree College and Alba graduate business school and the Scientific Director of the American College of Greece Research Center (ACG-RC). Prior to these, he was the Dean of Athens Information Technology (AIT), in Athens, Greece, where he was also Head of the Broadband Wireless and Sensor Networks (B-WiSE) Research Group. He currently holds Adjunct Professorships at Aalborg University and at the University of Cyprus. He received the Diploma of Electrical Engineering from the National Technical University of Athens (NTUA) in 1991 and the Doctorate degree in Signal Processing (highest honors) from the Ecole Nationale Supérieure des Télécommunications (ENST), Paris, France, in 1995. He was a researcher at Institut Eurécom (1992-1995), Stanford University (1995-1997) and Bell Labs (as Member of Technical Staff from 1997-2001 and as Technical Manager from 2001-2006). He was also Adjunct Professor at Columbia University (2004-2005) and Carnegie Mellon University (2006-2011). He has published over 220 papers and 4 books and has received approx. 10000 citations for his work, with an h-index of 45. He has also made standards contributions and holds 12 patents. He was a member of the Steering Board of the Wireless World Research Forum (WWRF) from 2002-2006, a member and industrial liaison of the IEEE’s Signal Processing for Communications Technical Committee from 2003-2008 and a National Representative of Greece to the European Research Council’s IDEAS program from 2007-2008. He has served as member of the IEEE Communications Society’s Fellow Evaluation and Awards Committees, as well as an Associate Editor for various journals, including the IEEE Trans. on Signal Processing. He was also a member of the IEEE Signal Processing for Communications for 2 terms. He has contributed to the organization of several IEEE conferences, including, as General Chair, the IEEE CTW 2016 and the IEEE SPAWC 2018 workshops. He has acted as Technical Coordinator in several EU projects such as: CROWN in the area of cognitive radio; HIATUS in the area of interference alignment; HARP in the area of remote radio heads and ADEL in the area of licensed shared access. He was the Research Coordinator of the European Training Network project PAINLESS on the topic of energy autonomous infrastructure-less wireless networks (2018-2022) and is currently the Technical Coordinator of the EU CHIST-ERA project FIREMAN on the topic of predictive maintenance via machine type wireless communication systems. His distinctions include the Bell Labs President’s Award (2002), the IEEE Signal Processing Society’s Young Author Best Paper Award (2003), a Bell Labs Teamwork Award (2004), his recognition as a “Highly Cited Greek Scientist” (2011), two IEEE conference paper awards (2013, 2014) and a “Best Booth” Award at EUCNC (2016). He has also been shortlisted twice for the Bell Labs Prize (2014, 2019). He was a Distinguished Lecturer of the IEEE Communications Society for 2012-2013. He is a Fellow of IEEE since 2013 and Fellow of the European Alliance of Innovation (EAI) since 2019. From 2020-2022 he was a member of Greece’s sectorial scientific council of Engineering Sciences, which supports the country’s National Council for Research and Innovation. He was recently elected Director-at-Large of the IEEE Signal Processing Society for Region 8 (Europe, Middle East and Africa).

\end{IEEEbiography}

\begin{IEEEbiography}[{\includegraphics[width=1in,height=1.25in,clip,keepaspectratio]{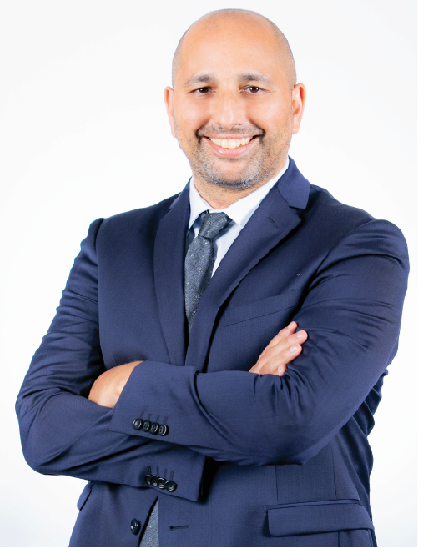}}]{Mérouane Debbah}

is Chief Researcher at the Technology Innovation Institute in Abu Dhabi. He is a Professor at Centralesupelec and an Adjunct Professor with the Department of Machine Learning at the Mohamed Bin Zayed University of Artificial Intelligence. He received the M.Sc. and Ph.D. degrees from the Ecole Normale Supérieure Paris-Saclay, France. He was with Motorola Labs, Saclay, France, from 1999 to 2002, and also with the Vienna Research Center for Telecommunications, Vienna, Austria, until 2003. From 2003 to 2007, he was an Assistant Professor with the Mobile Communications Department, Institut Eurecom, Sophia Antipolis, France. In 2007, he was appointed Full Professor at CentraleSupelec, Gif-sur-Yvette, France. From 2007 to 2014, he was the Director of the Alcatel-Lucent Chair on Flexible Radio. From 2014 to 2021, he was Vice-President of the Huawei France Research Center. He was jointly the director of the Mathematical and Algorithmic Sciences Lab as well as the director of the Lagrange Mathematical and Computing Research Center. Since 2021, he is leading the AI \& Digital Science Research centers at the Technology Innovation Institute. He has managed 8 EU projects and more than 24 national and international projects. His research interests lie in fundamental mathematics, algorithms, statistics, information, and communication sciences research. He is an IEEE Fellow, a WWRF Fellow, a Eurasip Fellow, an AAIA Fellow, an Institut Louis Bachelier Fellow and a Membre émérite SEE. He was a recipient of the ERC Grant MORE (Advanced Mathematical Tools for Complex Network Engineering) from 2012 to 2017. He was a recipient of the Mario Boella Award in 2005, the IEEE Glavieux Prize Award in 2011, the Qualcomm Innovation Prize Award in 2012, the 2019 IEEE Radio Communications Committee Technical Recognition Award and the 2020 SEE Blondel Medal. He received more than 20 best paper awards, among which the 2007 IEEE GLOBECOM Best Paper Award, the Wi-Opt 2009 Best Paper Award, the 2010 Newcom++ Best Paper Award, the WUN CogCom Best Paper 2012 and 2013 Award, the 2014 WCNC Best Paper Award, the 2015 ICC Best Paper Award, the 2015 IEEE Communications Society Leonard G. Abraham Prize, the 2015 IEEE Communications Society Fred W. Ellersick Prize, the 2016 IEEE Communications Society Best Tutorial Paper Award, the 2016 European Wireless Best Paper Award, the 2017 Eurasip Best Paper Award, the 2018 IEEE Marconi Prize Paper Award, the 2019 IEEE Communications Society Young Author Best Paper Award, the 2021 Eurasip Best Paper Award, the 2021 IEEE Marconi Prize Paper Award, the 2022 IEEE Communications Society Outstanding Paper Award, the 2022  ICC Best paper Award as well as the Valuetools 2007, Valuetools 2008, CrownCom 2009, Valuetools 2012, SAM 2014, and 2017 IEEE Sweden VT-COM-IT Joint Chapter best student paper awards. He is an Associate Editor-in-Chief of the journal Random Matrix: Theory and Applications. He was an Associate Area Editor and Senior Area Editor of the IEEE TRANSACTIONS ON SIGNAL PROCESSING from 2011 to 2013 and from 2013 to 2014, respectively. From 2021 to 2022, he served as an IEEE Signal Processing Society Distinguished Industry Speaker.

\end{IEEEbiography}

\end{document}